\documentclass[prd,onecolumn,superscriptaddress,12pt,nofootinbib]{revtex4-2}

\usepackage[]{graphicx}
\usepackage{dcolumn}
\usepackage{bm}

\usepackage{natbib}

\usepackage{amsmath,amssymb,amsfonts}
\usepackage{fancyhdr}
\usepackage{lipsum}
\usepackage{subcaption}
\usepackage{multirow}

\newcommand{\comment}[1]{}

\usepackage{color}
\usepackage[colorlinks]{hyperref}
\hypersetup{
    colorlinks = true,
    linkcolor = blue,
    anchorcolor = blue,
    citecolor = blue,
    filecolor = blue,
    urlcolor = blue
}

\usepackage{babel}
\newcommand{\be}{\begin{equation}}
\newcommand{\ee}{\end{equation}}
\newcommand{\bea}{\begin{eqnarray}}
\newcommand{\eea}{\end{eqnarray}}

\newcommand{\e}{\epsilon}

\newcommand{\half}{\frac{1}{2}}

\begin{document}

\title[]{\bf \Large The Simplest Oscillon and its Sphaleron\\ \vspace*{0.5cm}}

\author{\large \bf{N.~S. Manton}}
\email[]{N.S.Manton@damtp.cam.ac.uk}
\affiliation{Department of Applied Mathematics and Theoretical Physics,
University of Cambridge,
Wilberforce Road, Cambridge CB3 0WA, U.K.}

\author{\large \bf{T. Roma\'{n}czukiewicz}}
\email[]{tomasz.romanczukiewicz@uj.edu.pl}
\affiliation{Institute of Theoretical Physics, Jagiellonian University,
\L{}ojasiewicza 11, Krak\'{o}w, Poland}

\begin{abstract}

\vskip 10pt

Oscillons in a simple, 1-dimensional scalar field theory
with a cubic potential are discussed. The theory
has a classical sphaleron, whose decay generates a version of the
oscillon. A good approximation to the small-amplitude oscillon is
constructed explicitly using the asymptotic expansion of Fodor et
al., but for larger amplitudes a better approximation uses the discrete,
unstable and stable deformation modes of the sphaleron.
\end{abstract}
\maketitle

%%%%%%%%%%%%%%%%%%%%%%%%%%%%%
\section{Introduction}
%%%%%%%%%%%%%%%%%%%%%%%%%%%%%

Oscillons are spatially-localised, long-lived, oscillatory solutions
of the field equation(s) of classical field theories \cite{GleSic}. The
nonlinearity of the field equation is essential. Oscillons, unlike
kinks and other types of classical soliton, have no topological charge
ensuring their stability, and it is surprising that oscillons
do not couple more strongly to radiation modes of the field, leading to
rapid decay towards the classical vacuum.

Despite oscillons being known in a variety of field theories
in various spatial dimensions, the fundamental reason for their existence
remains somewhat mysterious. We will show that, at least for the
special, simple oscillon that we consider here in detail,
the oscillon can be thought of as a decaying sphaleron of the field theory.
By a sphaleron, we mean a localised, static but unstable solution of
the field equation \cite{KliMan}.

For an oscillon to exist, the continuum of radiation modes of the
linearised field needs to have a frequency gap, starting at some
positive threshold frequency $m$. A basic oscillon is periodic, with a
fundamental frequency $\omega < m$, so it couples to radiation only through
nonlinear terms, at frequencies that are multiples of $\omega$.
The oscillon has an arbitrary amplitude lying in some finite range
upwards from zero, and as the amplitude increases, the frequency
$\omega$ decreases away from the threshold $m$. Generally, $2\omega$ and higher
integer multiples of $\omega$ are in the continuum (although some exceptions are known \cite{Dorey:2019uap}), which underlines
the surprise that the oscillon is so long-lived. Nevertheless, an
oscillon does slowly radiate energy away, and as it does so its
amplitude decreases and its frequency increases. 

Much of the understanding of oscillons comes from numerical
investigation. One prototype is the oscillon in $\phi^4$ scalar field
theory, where the field potential is of the familiar double-well
form. This oscillon exists in the theory in 1-, 2- or 3-dimensions,
with the field profile depending just on radius and time (up to a
spatial translation). In the 1-dimensional theory, the oscillon is
reflection-symmetric about the origin. An oscillon of this type is
produced by starting from generic initial conditions of the form of a
symmetric hump, for example a Gaussian shape, superimposed on one of
the vacua. Oscillon formation is rather robust, and the initial shape
is not very important. There is usually a
transient in which the field shape changes over one
or two oscillations, with pulses of energy being radiated to the left
and right, and then the field settles into the oscillon.

A substantial theoretical analysis of oscillon structure was given by
Fodor et al. \cite{Fodor:2008es}, for oscillons of small and modest
amplitude \footnote{Refs. \cite{GleSic} and \cite{Fodor:2008es}
comprehensively review the oscillon literature up to 2009.}.
These authors considered a rather general scalar field
theory in spatial dimension $D \le 3$, whose potential $V(\phi)$ has a
Taylor expansion in $\phi$ about a quadratic minimum at $\phi = 0$. By an
iterative method, using the field equation, they systematically constructed
an oscillon as a series in an expansion parameter $\e$, related to the
amplitude. The oscillon's existence depends on the strength of the
cubic and higher-power terms in $V$, but the
conditions that arise are inequalities, so oscillons are
generic for small $\e$. By construction, the oscillon depends just
on radius and on time, and it is periodic (i.e. the Fourier series
w.r.t. time has terms that are multiples of a unique, fundamental
frequency $\omega$).

We shall use the method of Fodor et al. to explicitly construct a
particularly simple oscillon in 1-dimension. In practice, the
algebra is still quite tricky, and we have only found the first four
terms of the series in $\e$. As is hardly surprising, this series is
asymptotic rather than convergent, because
if the series were convergent then there would be a
strictly-periodic, exact oscillon solution, having infinite lifetime.
In practice, for small $\e$ it is useful to sum all four
terms to obtain a good approximation to the oscillon,
but as $\e$ and the amplitude increase, one needs to truncate the
series after fewer terms, as is typical for asymptotic series; the
discarded terms are larger than the last retained term.

For oscillons of even larger amplitude, the method of Fodor et al. tends to
break down, but instead, the oscillon can now be interpreted as arising
from the decay of a static sphaleron solution. The decaying sphaleron
can be well approximated using an ansatz constructed from two discrete
modes of the linearised deformations of the sphaleron, one unstable
and the other stable. This analysis shows that the sphaleron can be
regarded as the precursor of the oscillon.

The study of oscillons in field theories with double-well minima has
tended to hide this proposed connection between sphalerons and
oscillons. For example, the $\phi^4$ theory in 1-dimension with
double-well potential has no true sphaleron, but it has the configuration
of a kink and antikink at infinite separation as a `sphaleron'. If a
kink and antikink are released from a large separation at zero velocity,
and evolved numerically, then they turn into an oscillon (often called
a bion in this context). The sine-Gordon breather \cite{PerSky} provides
another example. This is an oscillon that lasts indefinitely because
of exact integrability. A large-amplitude breather instantaneously
comes to rest resembling a kink and antikink at large separation. Again,
the kink-antikink configuration at infinite separation can be thought
of as a sphaleron. The sine-Gordon breather exhibits a key
property of an oscillon, namely, that its fundamental frequency is
less than the continuum threshold for linearised waves, and as its
amplitude increases, the frequency decreases away from this threshold.

The connection between oscillons and sphalerons is clearer if there is
a genuine sphaleron of finite size in the field theory. Here, we
focus on a scalar field theory in 1-dimension which has such a
sphaleron. We assume that $V(\phi)$ has a quadratic minimum at
$\phi = 0$, with $V(0) = 0$. Then, a sphaleron exists if $V$ becomes
negative for some $\phi > 0$. (It is convenient to choose this sign
for the inequality, but $\phi < 0$ is equivalent.) More simply, we assume
that $V(\phi)$ increases to a local maximum at $\phi = \phi_1 > 0$,
then decreases and passes linearly through $V=0$ at some
$\phi_2 > \phi_1$. $V$ could have further local or global minima
as $\phi$ increases further. Note that $\phi = 0$, which
is the asymptotic value of the sphaleron tail field, is a {\it false}
vacuum, because it is not the global minimum of $V$, but this
doesn't cause difficulties.

The existence of a time-independent sphaleron solution in 1-dimension
can be easily understood
using the standard trick of identifying the static field equation as the
equation for a particle rolling in the inverted potential $-V$. In
the inverted potential, the particle starts at rest from $\phi=0$, rolls
through the potential minimum at $\phi=\phi_1$ and ascends the potential to
$\phi_2$. As the potential is linear here, the particle stops
instantaneously, then rolls back to the starting point at
$\phi = 0$. Because $V$ is quadratic around $\phi=0$,
the whole process takes infinite time. Spatially, one
obtains a hump-shaped sphaleron profile which has field values lying
in the range $0 < \phi \le \phi_2$, with tails approaching $\phi = 0$
exponentially fast.

The connection between oscillons and sphalerons in a potential of this
type seems first to have been noted in ref.\cite{Dorey:2021mdh}, but
here we will explore the connection more systematically. We will work with the
simplest potential of the required form, the purely cubic potential
$V(\phi) = \half \phi^2 - \frac{1}{3} \phi^3$. $V$ has a quadratic
local minimum at $\phi = 0$, with value zero, a local maximum at
$\phi_1 = 1$, and passes through zero linearly at $\phi_2 = \frac{3}{2}$. The
sphaleron has a simple analytical form, and we can calculate its
unstable mode, its translation zero mode, and its single discrete
vibrational mode -- its shape mode. At the same time, this potential
allows for explicit calculation of a small-amplitude oscillon as a series,
using the method of Fodor et al., and we have analytically calculated the
terms up to fourth order in the expansion parameter $\e$.
Importantly, we will show numerically that if the
sphaleron is perturbed by its unstable mode, in the direction of
decreasing $\phi$, then it evolves into the oscillon. (If
it is perturbed in the opposite direction, then the field values
quickly become very large, and the field becomes singular.) 

This sphaleron in 1-dimension, arising from a cubic potential, is
not new. It occurs as a ``bounce'' in work of Callan and
Coleman \cite{CallanCurtis:1977} and
was discussed in detail by Avelar et al. \cite{Avelar:2008}. These
authors noted that because its translation zero mode has a node,
there must be a mode with negative squared frequency, i.e. an
instability. Avelar et al. also found the positive-frequency shape
mode. However, the connection to oscillons appears not to have
been investigated before.

Clearly, at the linearised level, the instability and shape
oscillation of the sphaleron can be modelled by combining the
sphaleron with the two relevant discrete
modes. As we shall only consider the sphaleron and
oscillon with their centres of mass at rest, we can ignore the
translation zero mode. The perturbed sphaleron, like the oscillon,
is then reflection-symmetric. We shall now make a bold leap, and consider
the sphaleron deformed by these two modes with arbitrarily large,
time-dependent amplitudes. This is a collective coordinate ansatz
for the evolution of the sphaleron. The reduced, collective coordinate
Lagrangian, obtained by substituting this ansatz into the
field-theoretic Lagrangian, is nonlinear but rather simple. We will
show that its resulting dynamics gives another good approximation to the
oscillon, which is particularly useful when the oscillon has quite large
amplitude and the series of Fodor et al. breaks down.

We should clarify here that the oscillon constructed by the method of
Fodor et al. has just one degree of freedom, its amplitude, and
its shape and frequency depend on this. Numerically however, one
typically finds that an oscillon appears to be quasi-periodic,
although by careful adjustment of initial conditions, the periodic
version can be found too. To model quasi-periodic behaviour one needs to
have a system with two degrees of freedom at least, and the two
mode amplitudes of the deformed sphaleron provide these. This issue was also
recently raised by Blaschke and Karp\'\i{}\v{s}ek \cite{Blaschke:2022fxp},
who studied a mechanised model of an oscillon 
with two internal degrees of freedom (in addition to the centre of
mass position). In fact, a non-integrable Lagrangian system with two degrees
of freedom (and 4-dimensional phase space) has more complicated
dynamics than quasi-periodic motion, but we have not been able to
observe this in the oscillon. The issue of quasi-periodic or chaotic
behaviour of the oscillon is complicated, because the
reduced system is only an approximation to the field theory with
its infinitely many degrees of freedom, and does not couple to radiation.

It is surprising that a model using the sphaleron's two linearised modes
is quantitatively useful, because there are no values of the two mode
amplitudes giving the vacuum field configuration $\phi \equiv 0$
exactly (although, for optimal values, it gets quite close).
Consequently, this rather crude model cannot accurately describe oscillons of
small amplitude.  

In the following, we introduce the scalar field theory with cubic
potential, then construct the first four terms in the series for the
small-amplitude oscillon solution, following Fodor et al. Next, we recall the
sphaleron solution and its discrete modes, and use these to
construct and test our collective coordinate dynamics modelling an
oscillon of larger amplitude. Finally, we describe in further detail
some features of the oscillon that we have uncovered numerically, and
present our conclusions.  

\section{A Simple Scalar Field Theory}

\begin{figure}
 \includegraphics[width=0.75\columnwidth]{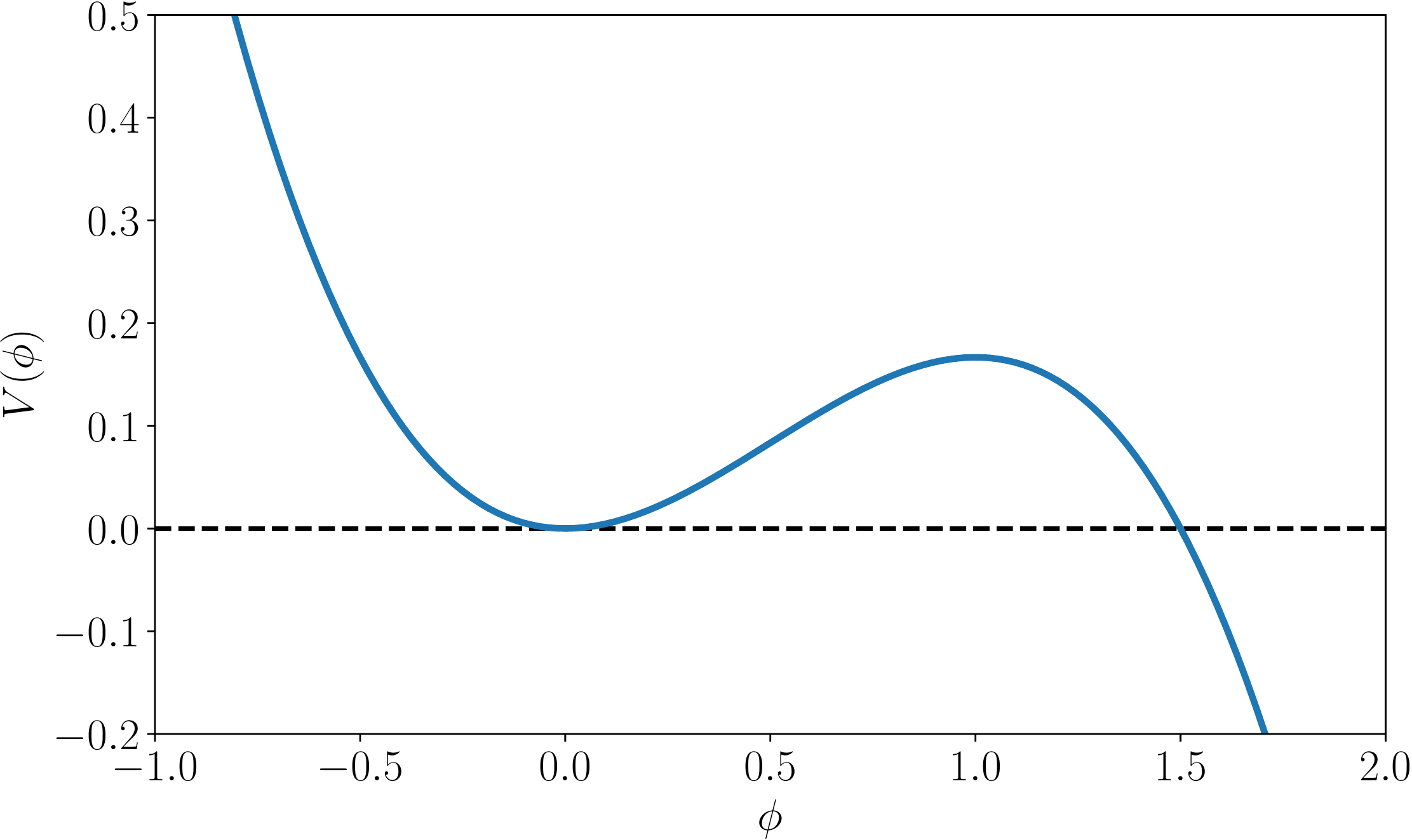}
 \caption{Potential $V(\phi) = \half \phi^2 - \frac{1}{3} \phi^3$.}
 \label{fig:potential}
\end{figure}

Consider the theory for a real scalar field $\phi(t,x)$ in 1-dimension,
with Lagrangian
\be
L[\phi] = \int_{-\infty}^\infty \left(\half\phi_t^2-\half\phi_x^2
-\half\phi^2+\frac{1}{3}\phi^3\right)dx \,.
\label{Lagran}
\ee
This has the simple nonlinear field equation
\be
\phi_{tt} - \phi_{xx} + \phi - \phi^2 = 0 \,.
\label{fieldeq}
\ee
FIG. \ref{fig:potential} shows the potential
\be
V(\phi) = \half\phi^2-\frac{1}{3}\phi^3 \,,
\label{PotV}
\ee
which is unbounded below but has a local quadratic minimum at
$\phi = 0$ with $V(0)$ zero, and a local maximum at $\phi = 1$ with 
$V(1)=\frac{1}{6}$. Additionally, $V$ passes linearly through zero
at $\phi = \frac{3}{2}$.

\section{The Small-Amplitude Oscillon}

Following the method of Fodor et al. \cite{Fodor:2008es} to construct
an oscillon solution of eq.(\ref{fieldeq}), we expand
the field in powers of a small parameter $\e$,
\be
\phi = \sum_{k=1}^\infty \e^k\phi_k(t,x) \,.
\label{oscexpan}
\ee
We denote the truncated series as $\Phi_N=\sum_{k=1}^N \e^k\phi_k(t,x)$.
We also introduce rescaled space and time variables
\be
\zeta = \e x, \qquad \tau=\omega t
\ee
and assume that
\be
\omega=\sqrt{1-\e^2} \,,
\ee
which locks the expansion parameter $\e$ to the oscillon frequency. We
assume the oscillon is instantaneously at rest at $\tau = 0$, so
it is symmetric in $\tau$. The oscillon will also be symmetric in
$\zeta$, and we can identify its amplitude as $ \sum_{k=1}^\infty
\e^k \phi_k(0,0)$, or the truncated version of this. In terms of
these new variables the field equation (\ref{fieldeq}) takes the form
\be
(1-\e^2)\ddot\phi - \e^2\phi'' + \phi - \phi^2 = 0 \,,
\ee
where overdots and primes denote derivatives w.r.t. $\tau$ and $\zeta$
respectively. Expanding in powers of $\e$, we obtain an infinite
set of coupled equations, of which the first five are
\bea
\ddot{\phi_1} + \phi_1 &=& 0 \,, \label{phi1} \\
\ddot{\phi_2} + \phi_2 &=& \phi_1^2 \,, \label{phi2} \\
\ddot{\phi_3} + \phi_3 &=& \ddot{\phi_1} + \phi_1'' + 2\phi_1\phi_2 
\,, \label{phi3} \\
\ddot{\phi_4} + \phi_4 &=& \ddot{\phi_2} + \phi_2'' + 2\phi_1\phi_3
+ \phi_2^2 \,, \label{phi4} \\
\ddot{\phi_5} + \phi_5 &=& \ddot{\phi_3} + \phi_3'' + 2\phi_1\phi_4
+ 2\phi_2\phi_3 \,. \label{phi5}
\eea
These can be regarded as an iterative sequence of ordinary, linear
differential equations for $\phi_1, \phi_2, \phi_3, \dots$, whose
sources on the right-hand side depend on the previously determined
functions. [Note that in ref.\cite{Fodor:2008es}, eqs.(\ref{phi4})
and (\ref{phi5}) are not given explicitly, and their version of
eq.(\ref{phi3}) has a typo; their explicit $-\ddot{\phi_1}$
should be left out, as it is present in the term $\omega_2\ddot{\phi_1}$.]

The solution of eq.(\ref{phi1}), symmetric in $\tau$, is
\be
\phi_1 = p_1(\zeta) \cos \tau
\ee
where $p_1$ is yet to be determined. Equation (\ref{phi2}) now becomes
$\ddot{\phi_2} + \phi_2 = \half p_1^2(\zeta)(1 + \cos 2\tau)$,
whose solution, combining the particular integral with a homogeneous
function symmetric in $\tau$, is
\be
\phi_2 = p_2(\zeta) \cos \tau + \frac{1}{6} p_1^2(\zeta)
(3 -\cos 2\tau) \,.
\ee

The two unknown functions, $p_1$ and $p_2$, are determined by considering
eqs.(\ref{phi3}) and (\ref{phi4}) for $\phi_3$ and $\phi_4$. First, for the
oscillon to be periodic, there is a condition of {\it no resonance},
i.e. the right-hand side of eq.(\ref{phi3}) should have no $\cos \tau$
term. This condition reduces to
\be
p_1'' - p_1 + \frac{5}{6} p_1^3 = 0 \,,
\ee
whose solution symmetric in $\zeta$, and decaying for large $|\zeta|$, is
\be
p_1(\zeta) = \sqrt{\frac{12}{5}} \frac{1}{\cosh \zeta} \,.
\ee
Second, one finds that it is consistent to set $p_2 \equiv 0$. This can be
argued in more than one way. After solving for $\phi_3$,
it is found that the no resonance condition for
$\phi_4$ implies that $p_2$ obeys a linear differential equation
whose solution is an antisymmetric function of $\zeta$, whereas we
require the oscillon to be symmetric in $\zeta$. Setting $p_2 \equiv 0$
also means that the oscillon can be symmetric under the combined
transformations $\e \to -\e$, $\tau \to \tau + \pi$. More generally,
the latter symmetry requires that $\phi_k$ only has terms $\cos n\tau$ with
$n$ even/odd when $k$ is even/odd. In summary, we have established
that the leading terms in the series for the oscillon are
\be
\phi_1 = \sqrt{\frac{12}{5}} \frac{\cos \tau}{\cosh \zeta} \,,
\quad \phi_2 = \frac{2}{5} \, \frac{3 - \cos 2\tau}{\cosh^2\zeta} \,.
\label{phi1phi2}
\ee

Equation (\ref{phi3}) now simplifies, and its solution is
\be
\phi_3 = p_3(\zeta)\cos \tau + \sqrt{\frac{12}{5}} \frac{1}{20}
\frac{\cos 3\tau }{\cosh^3\zeta} \,,
\label{phi3soln}
\ee
where $p_3$, the homogeneous contribution, is as yet arbitrary and will
not be zero. It is then straightforward to substitute for
$\phi_1, \phi_2$ and $\phi_3$ in eq.(\ref{phi4}), and
integrate to find that
\be
\phi_4 = \sqrt{\frac{12}{5}}\frac{1}{3}\frac{p_3(\zeta)}{\cosh \zeta} 
(3 -\cos 2\tau) + \frac{24}{5}\frac{1}{\cosh^2\zeta}
- \frac{1}{75}\frac{426 + 39\cos 2\tau +\cos 4\tau}{\cosh^4\zeta} \,.
\label{phi4soln}
\ee
There could be an additional homogeneous term $p_4(\zeta)\cos \tau$,
but the symmetry mentioned above requires $\phi_4$ only to have terms
proportional to $\cos n\tau$ with $n$ even, so we can set $p_4 \equiv 0$. 

Finally, we impose the no resonance condition for $\phi_5$, i.e. that
there is no $\cos \tau$ term on the right-hand side of
eq.(\ref{phi5}). This gives an ordinary differential equation for
$p_3$, of the P\"oschl--Teller form with a source, whose
acceptable solution is
\be
p_3(\zeta) = \sqrt{\frac{12}{5}} \frac{1}{60}\left( \frac{94}{\cosh \zeta}
- \frac{119}{\cosh^3 \zeta} \right) \,.
\ee
Combined with the earlier results (\ref{phi3soln}) and (\ref{phi4soln}),
this gives the final form for $\phi_3$ and $\phi_4$,
\bea
\phi_3 &=& \sqrt{\frac{12}{5}}\frac{1}{60}
\left( \frac{94\cos \tau}{\cosh \zeta}
- \frac{119\cos \tau - 3\cos 3\tau}{\cosh^3\zeta} \right) \,, \nonumber \\
\phi_4 &=& \frac{1}{75} \left( \frac{642 - 94\cos 2\tau}{\cosh^2\zeta}
- \frac{783 - 80\cos 2\tau + \cos 4\tau}{\cosh^4\zeta} \right) \,.
\label{phi3phi4}
\eea
We do not calculate $\phi_5$ as this will involve yet another
non-zero arbitrary function $p_5$ that can only be determined by a no
resonance condition in the equation for $\phi_7$.

\begin{figure}
 \includegraphics[width=1\columnwidth]{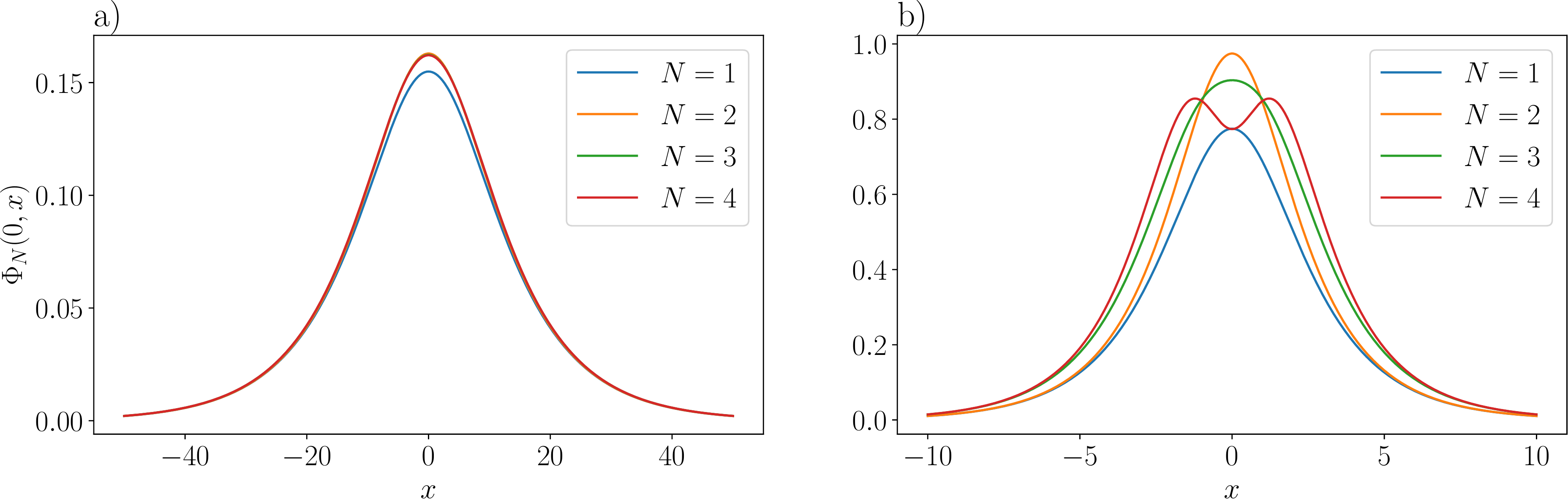}
 \caption{Profiles of the oscillons at $\tau=0$ for truncation orders
   $N=1, \dots, 4$ of the Fodor et al. series -- a) $\epsilon=0.1$
   and b) $\epsilon=0.5$. $x$ is the unscaled spatial variable.}
 \label{fig:profile_approx1}
\end{figure}

\begin{figure}
 \includegraphics[width=1\columnwidth]{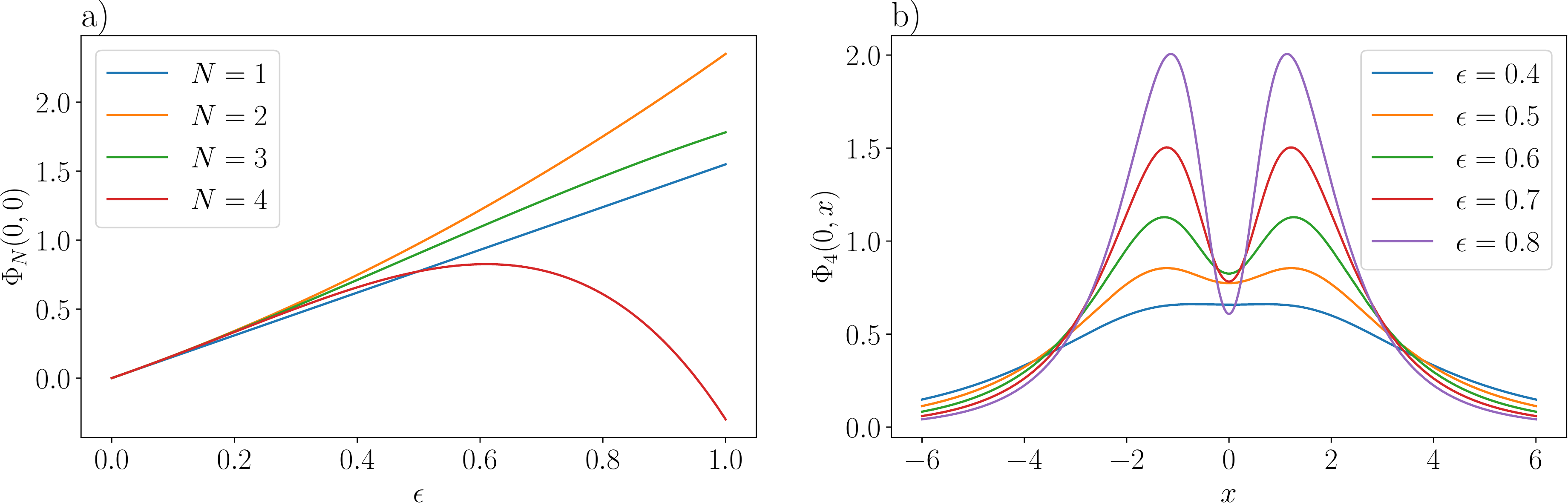}
 \caption{a) Value of the field profile $\Phi_N$ at the center
   $\zeta=0, \tau=0$ for varying $\e$. b) Profile of the truncated
   oscillon $\Phi_4$ for larger values of $\e$.}
 \label{fig:profile_approx}
\end{figure}

The truncated approximate oscillon, $\Phi_N$,
is the sum of the first $N$ terms of the expansion (\ref{oscexpan}), where
$\phi_1,\dots,\phi_4$ are as in eqs.(\ref{phi1phi2}) and (\ref{phi3phi4}).
FIGS. \ref{fig:profile_approx1} show this truncated oscillon at
$\tau = 0$ for $N = 1, \dots, 4$, and for amplitude parameters
$\e = 0.1$ and $\e = 0.5$. FIG. \ref{fig:profile_approx} a)
shows the combined strength of the contributing terms,
evaluated at $\zeta = \tau = 0$. It is clear that for $\e \gtrsim 0.6$,
the higher-order terms are no longer small compared to the lower-order
terms, as is typical for an asymptotic series, so it is better to
truncate the series after two or three terms, obtaining $\Phi_2$ or
$\Phi_3$. The pronounced double-hump of the oscillon profile for
large $\e$ in FIG. \ref{fig:profile_approx} b) appears therefore to
be exaggerated, and not a reliable feature.

The truncated oscillon has just one degree of freedom, $\e$, and it
is periodic with $t$-period $2\pi/\sqrt{1 - \e^2}$. This is because of
the symmetry assumptions that have been imposed. There were
opportunities to include less symmetric terms in the construction,
so a larger family of oscillons could probably be found, although
more algebraic work would be required. There is therefore no
inconsistency with the approach discussed below, where the oscillon
is generally quasi-periodic.

\begin{figure}
 \includegraphics[width=0.75\columnwidth]{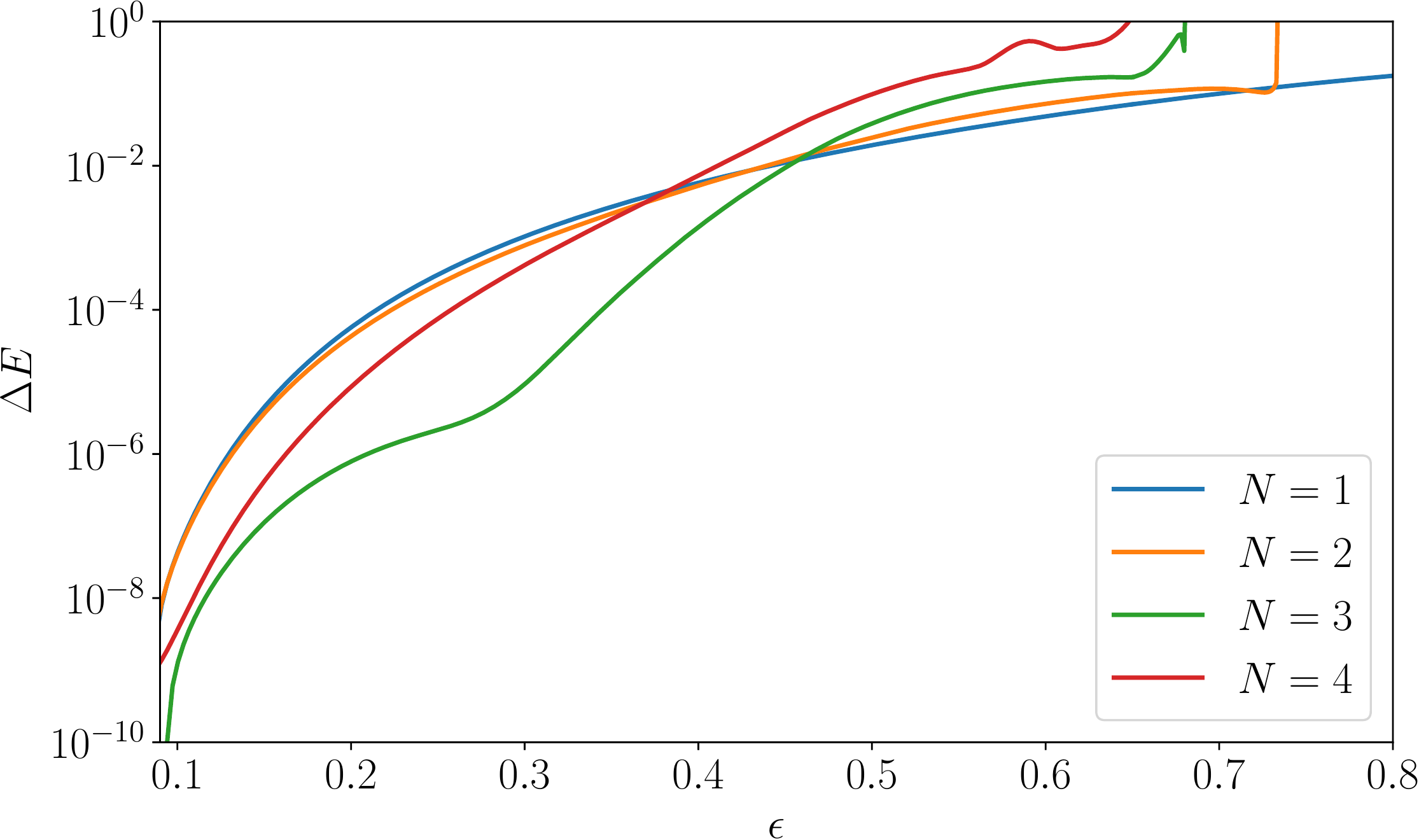}
 \caption{Oscillon energy loss from the spatial interval $[-100, 100]$
   during time $0<t<300$, starting from the truncated series
   $\Phi_N(0,x)$ as initial configuration.}
 \label{fig:flux}
\end{figure}

To show the quality of the truncated oscillon, we have
numerically solved the field equation (using variables $t,x$) with
initial condition $\Phi_N(0, x)$ for $N=1, \dots, 4$ and a wide range of
$\e\in[0.1,0.8]$. We have measured the loss of energy from the spatial
interval $-100 < x < 100$ during the time interval $0<t<T=300$. The
energy loss $\Delta E$ is the time-integrated energy flux
through the ends, which is equal on the left and right, so 
\begin{equation}
\Delta E = 2\int_0^T \phi_t(t,100)\phi_x(t,100) \, dt \,,
\label{flux}
\end{equation}
and is shown in FIG. \ref{fig:flux}. In the range $\e\in[0.1, 0.5]$,
$\Phi_3$ is the best initial condition. For larger $\epsilon$,
the initial configuration $\Phi_4$ loses energy faster, and the
approximate oscillon $\Phi_4(t,x)$ breaks down. $\Phi_4$
is probably a better approximation to the numerical solution than
$\Phi_3$ for small values of $\epsilon$, but this is not clear from
the figure because of possible numerical errors.

\section{The Sphaleron}

There exists a nontrivial, lump-like static solution of the field
equation (\ref{fieldeq}),
\be
\phi_{\rm S}(x) = \frac{3}{2}\frac{1}{\cosh^2 \half x} \,.
\label{sphal}
\ee
This is expressed in terms of the unscaled spatial variable $x$ and
satisfies the boundary conditions $\phi_{\rm S} \to 0$ as $x \to \pm
\infty$, like the oscillon. The solution can be
translated, but as given, it is reflection-symmetric in $x$. 
Its energy is $E = \frac{6}{5}$.

A small perturbation $\delta\phi=e^{i\omega t}\eta(x)$
of $\phi_{\rm S}$, with frequency $\omega$, obeys the 
linearised equation
\be
-\eta''(x)+U(x)\eta(x) = \omega^2\eta(x) \,,
\label{sphalpert}
\ee
where
\be
U(x) = V''(\phi_{\rm S}(x)) = 1 - \frac{3}{\cosh^2 \half x} \,.
\ee
$U$ is a P\"oschl--Teller potential, so the solutions of
eq.(\ref{sphalpert}) are well known. There are three (normalised)
discrete modes,
\bea
\eta_{-1}(x) = \sqrt{\frac{15}{32}} \, \frac{1}{\cosh^3 \half x} \,,
&& \qquad \omega_{-1}^2 = -\frac{5}{4} \,, \\
\eta_0(x) = \sqrt{\frac{15}{8}} \, \frac{\sinh \half x}{\cosh^3 \half x} \,,
&& \qquad \omega_0^2 = 0 \,, \\
\eta_1(x) = \sqrt{\frac{3}{32}} \, \frac{4\cosh^2 \half x \, - \, 5}
{\cosh^3 \half x} \,, && \qquad \omega_1^2 = \frac{3}{4} \,.
\eea

The presence of a unique unstable mode $\eta_{-1}$ with negative squared
frequency means that the lump is a sphaleron. It is the saddle point in
field configuration space between the false vacuum $\phi \equiv 0$
(with zero energy) and configurations with negative energy, whose field
$\phi$ is large and positive in some region of physical space. After
being perturbed in the unstable direction towards the false vacuum,
the sphaleron's evolution connects it with the oscillon. The
sphaleron's discrete shape mode $\eta_1$, whose positive frequency
$\omega_1$ is below the continuum threshold at $\omega = 1$, is also
important. It is the source of a second degree of freedom for the
oscillon. $\eta_0$ is the translation zero mode, and can be ignored
here, because it has the opposite reflection symmetry to the other
modes, and doesn't contribute to an oscillon whose centre of mass is at rest.

\begin{figure}
 \includegraphics[width=0.75\columnwidth]{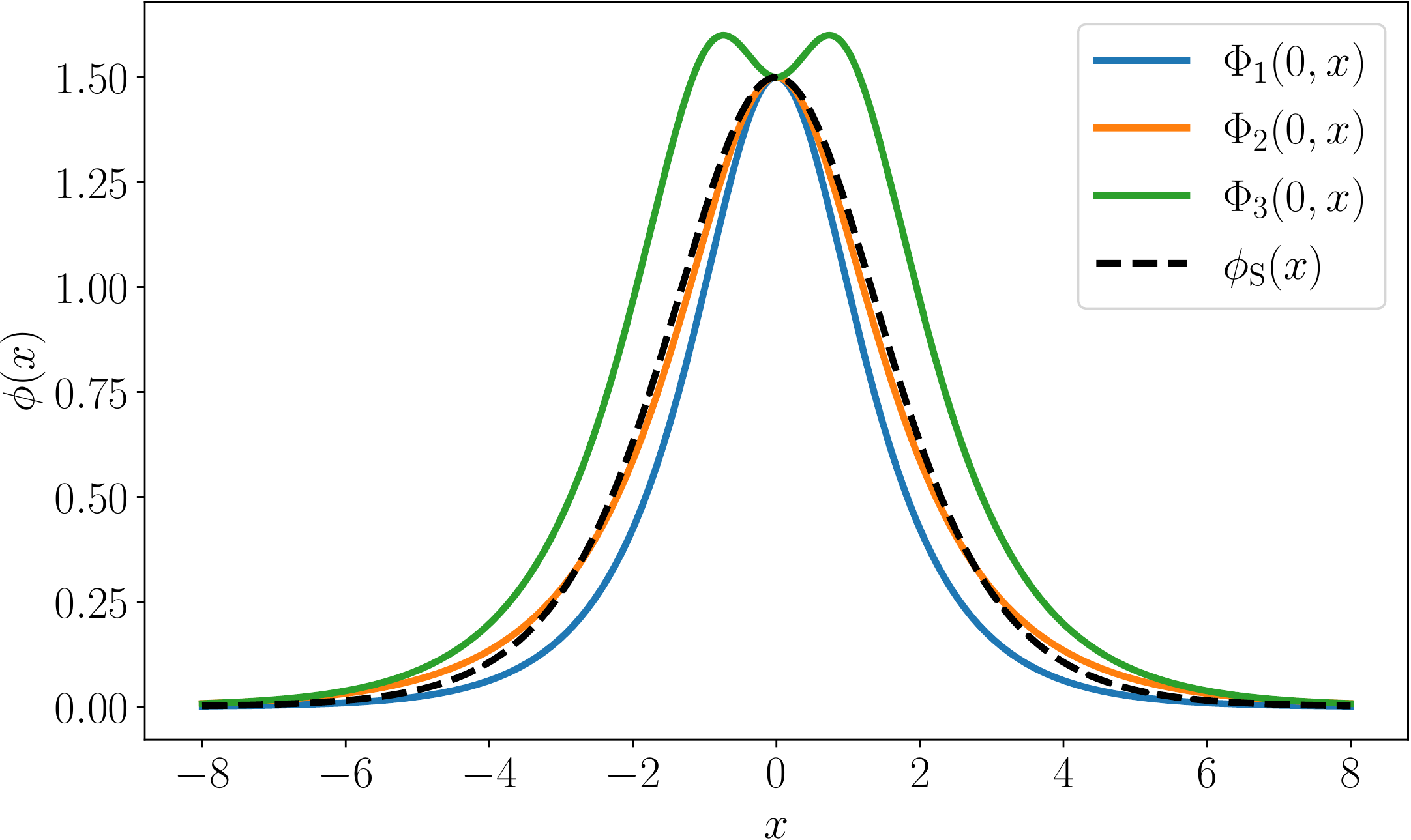}
 \caption{Match between the sphaleron $\phi_{\rm S}(x)$ and the
   profiles $\Phi_N(0, x)$ for optimal $\epsilon$.}
 \label{fig:sphaleron_approx}
\end{figure}

It is instructive to compare the truncated oscillon profiles
$\Phi_N(0, x)$ for varying $\e$ with the sphaleron profile
(FIG. \ref{fig:sphaleron_approx}). $\Phi_1(0, 0)$ matches the
sphaleron central amplitude $\phi_{\rm S}(0)=\frac{3}{2}$ for
$\epsilon=\frac{3}{2}\sqrt{\frac{5}{12}}\approx0.9682$.
This corresponds to an oscillon frequency $\omega=\frac{1}{4}$.
However, the $L^2$ norm $\|\phi_{\rm S}(x)-\Phi_1(0,x)\|^2\approx0.1776$
shows that the match of profiles is not good. The second truncation
matches much better. The condition $\Phi_2(0, 0)=\frac{3}{2}$
is a quadratic equation with solutions
\begin{equation}
  \epsilon_1=\frac{15-5\sqrt{3}}{4\sqrt{5}}\approx0.7088 \,,
  \qquad\epsilon_2=\frac{-15-5\sqrt{3}}{4\sqrt{5}}\approx-2.6453 \,.
\end{equation}
The second solution is outside the acceptable range, $-1<\epsilon<1$, but
the first gives a profile very close to the sphaleron with
$\|\phi_{\rm S}(x)-\Phi_2(0,x)\|^2\approx0.0138$.
The corresponding oscillon frequency is $\omega\approx0.7054$.
$\Phi_3$ has a profile with a dip for large $\e$ and matches much
worse, as does $\Phi_4$.

\section{Collective Coordinate Model based on the Sphaleron}

Reflection-symmetric field evolution around the sphaleron,
including the (normalised) unstable and shape modes
$\eta_{-1}$ and $\eta_1$, can be modelled by the ansatz
\be
\phi(t,x) = \phi_{\rm S}(x) + A(t) \, \eta_{-1}(x) + B(t) \, \eta_1(x) \,.
\label{ansatz}
\ee
At the linearised level, $B$ oscillates and $A$ tends to grow
exponentially, suggesting that the ansatz will be valid for limited
time. Rather surprisingly, this ansatz has an extended approximate
validity. $A$ and $B$ can be assumed to have unconstrained magnitudes,
and can be treated as collective coordinates of the field $\phi$.
Their nonlinear time-evolution gives an approximate model for the oscillon. To
find the model equations, we substitute the ansatz (\ref{ansatz})
into the field Lagrangian (\ref{Lagran}). After evaluating the
derivatives and integrating over space (and discarding boundary terms),
we obtain a reduced, effective Lagrangian whose nonlinear equations
of motion define the collective coordinate dynamics.

Because the discrete modes are localised, they
provide a useful approximation to the oscillon. This is especially true
for oscillons whose amplitude is not too small. Recall that an oscillon of
small amplitude has a large spatial extent (since in the Fodor et
al. analysis it is a function of the scaled spatial variable
$\zeta = \e x$). Oscillons of larger amplitude have a shape closer to that
obtained by deforming the sphaleron by its discrete modes.

The reduced Lagrangian is of the form
\be
L_\text{eff}[A,B] = \half\dot{A}^2 + \half\dot{B}^2 - V_\text{eff}(A, B) \,,
\label{Lageff}
\ee
where overdots are now unscaled time-derivatives and
\be
V_\text{eff}(A,B) = \frac{6}{5} - \frac{5}{8}A^2 + \frac{3}{8}B^2 -
C_1 A^3 - C_2 A^2 B - C_3 A B^2 - C_4 B^3 \,,
\ee
with the constants $C_1, \dots, C_4$ given below.
To establish this, some integration by parts is needed, together with use
of the nonlinear equation satisfied by $\phi_{\rm S}$ and the
linearised equations for the retained modes. The kinetic terms have
a simple Euclidean form because the modes are orthonormal. The first
three coefficients in the potential $V_\text{eff}$ are the energy of
the sphaleron and half the (negative and positive) squared frequencies
of the retained modes. The coefficients of the cubic terms are the integrals 
\be
C_1 = \frac{1}{3} \int_{-\infty}^\infty \eta_{-1}^3(x) \, dx =
\sqrt{\frac{15}{2}}\frac{175\pi}{8192} \,,
\quad C_2 = \int_{-\infty}^\infty \eta_{-1}^2(x)\eta_1(x) \, dx =
-\sqrt{\frac{3}{2}}\frac{225\pi}{8192} \,, \nonumber
\ee
\be
C_3 = \int_{-\infty}^\infty \eta_{-1}(x)\eta_1^2(x) \, dx =
\sqrt{\frac{15}{2}}\frac{129\pi}{8192} \,,
\quad C_4 = \frac{1}{3} \int_{-\infty}^\infty \eta_1^3(x) \, dx =
\sqrt{\frac{3}{2}}\frac{201\pi}{8192} \,.
\ee

From the reduced Lagrangian (\ref{Lageff}) we obtain the equations of motion
\begin{subequations}
\bea
\frac{d^2 A}{dt^2} &=& \frac{5}{4} A + 3C_1 A^2 + 2C_2 AB + C_3 B^2 \,,
\label{EffectiveEqsa} \\
\frac{d^2 B}{dt^2} &=& -\frac{3}{4} B + C_2 A^2 + 2C_3 AB + 3C_4 B^2 \,.
\label{EffectiveEqsb}
\eea
\end{subequations}
and the conserved energy
\be
E_\text{eff}[A,B] = \half\dot{A}^2 + \half\dot{B}^2 + V_\text{eff}(A,B) \,,
\label{Eeff}
\ee

\begin{figure}
 \includegraphics[width=0.75\columnwidth]{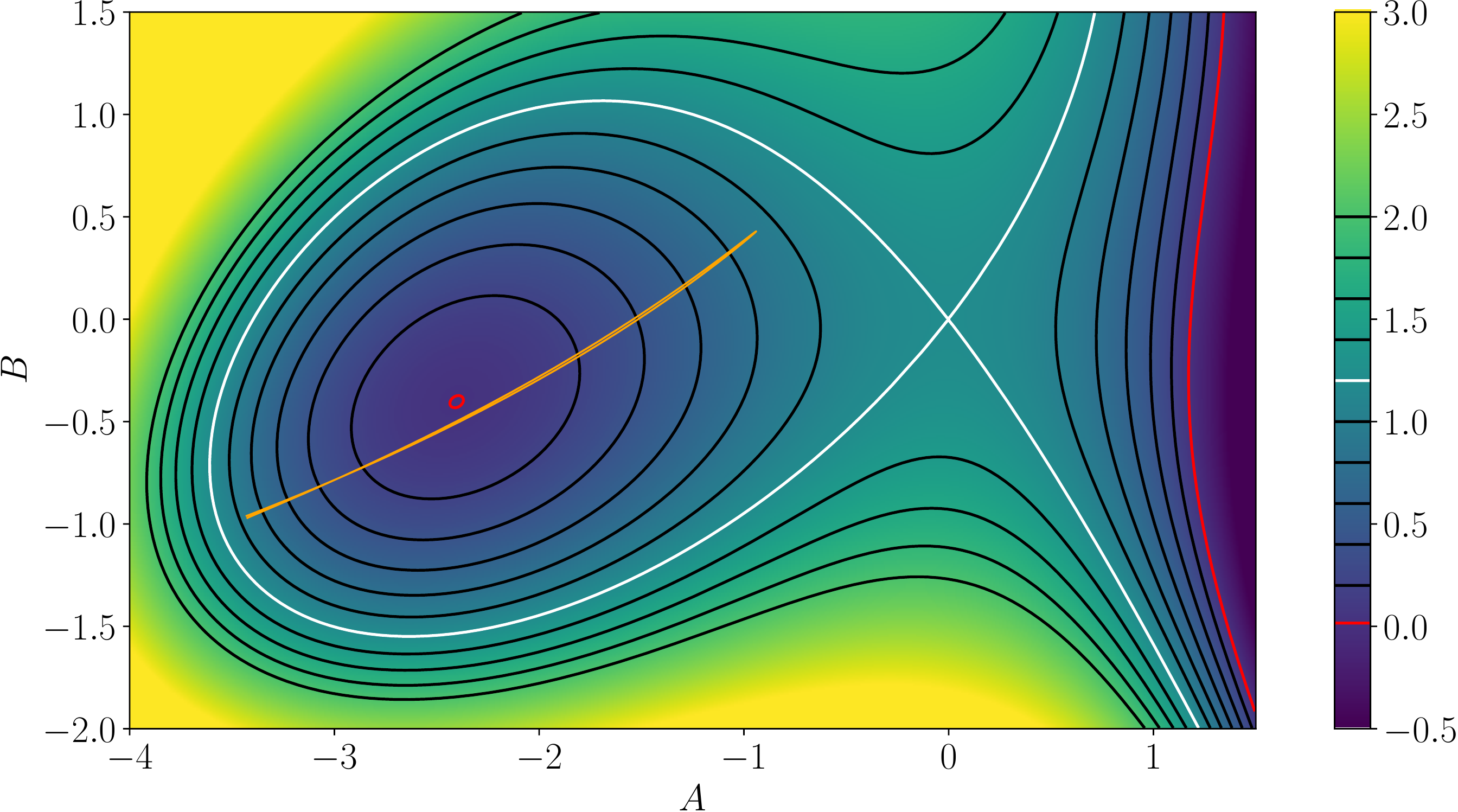}
 \caption{Effective potential $V_\text{eff}(A, B)$. The white contour
   corresponds to the energy 1.2 of the sphaleron and the pair of
   red contours to energy $0.016$, slightly above that of the approximate
   (false) vacuum. The orange line is a trajectory of a solution discussed in
   section \ref{sec:comparison} and FIG. \ref{fig:comparison_good}.}
 \label{fig:effective_potential}
 \end{figure}

\begin{figure}
 \includegraphics[width=0.75\columnwidth]{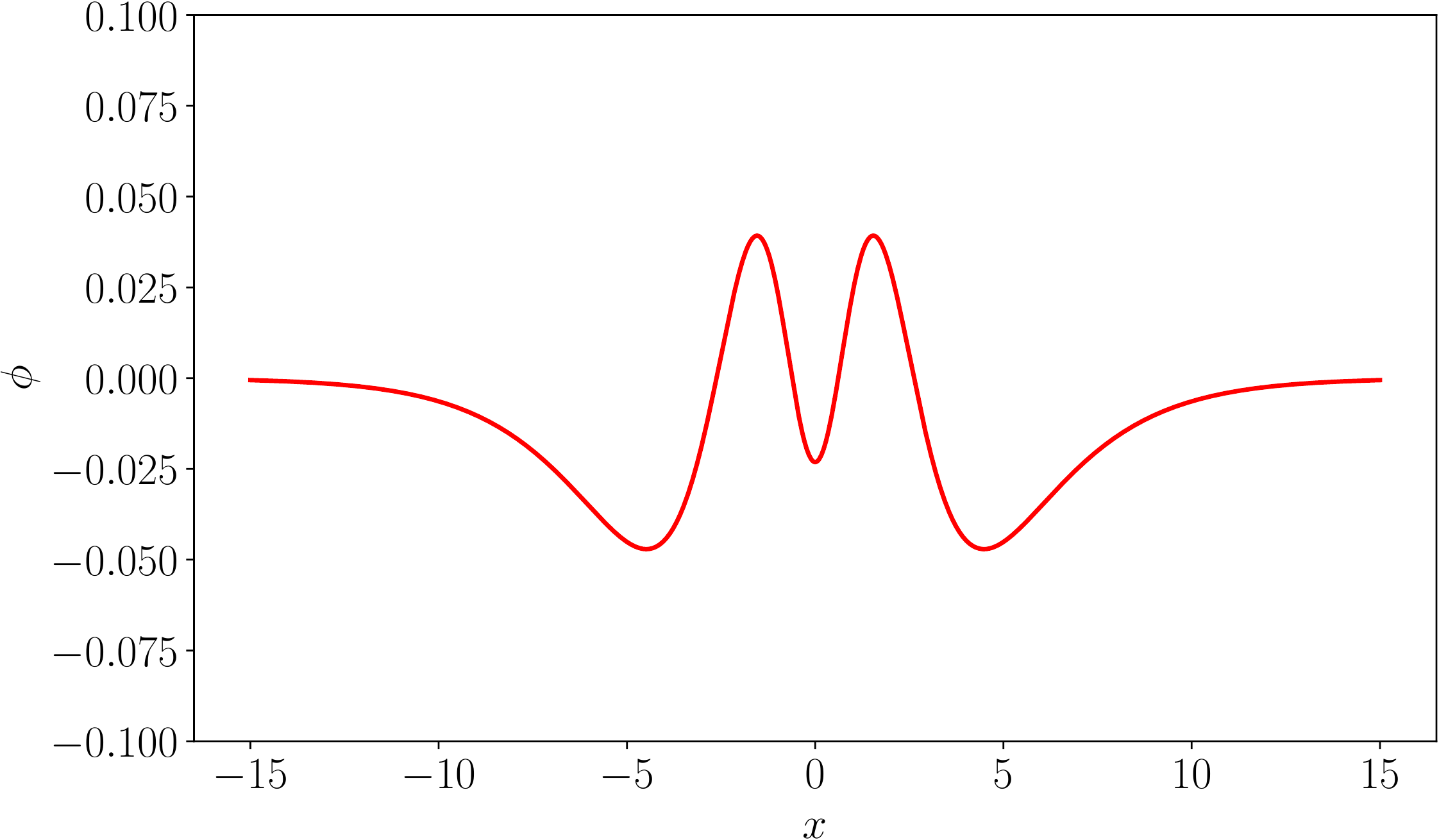}
 \caption{Optimal approximation to the (false) vacuum configuration
   using the sphaleron plus modes expansion (\ref{ansatz}).}
 \label{fig:approx_vacuum}
\end{figure}

FIG. \ref{fig:effective_potential} shows a contour plot of $V_\text{eff}$.
There is a saddle point at $A = B = 0$ corresponding to the sphaleron,
and a local minimum at $A = -2.40501 \,, B = -0.40325$ corresponding
to an approximation to the (false) vacuum $\phi \equiv 0$, whose energy is
$0.01532$ and whose field configuration (\ref{ansatz}) is shown in
FIG. \ref{fig:approx_vacuum}. In the reduced dynamics, diagonalised small
perturbations around the approximate vacuum have frequencies
\mbox{$\tilde\omega_1 = 1.02216$} and $\tilde\omega_2 = 1.37920$, which
are above the continuum threshold frequency $\omega = 1$. This is partly
because the minimum is not the exact vacuum, but mainly
because the perturbations are linear combinations of the localised
modes $\eta_{-1}$ and $\eta_1$, which do not have the large wavelengths
of radiation modes close to the threshold.

\begin{figure}
 \includegraphics[width=0.85\columnwidth]{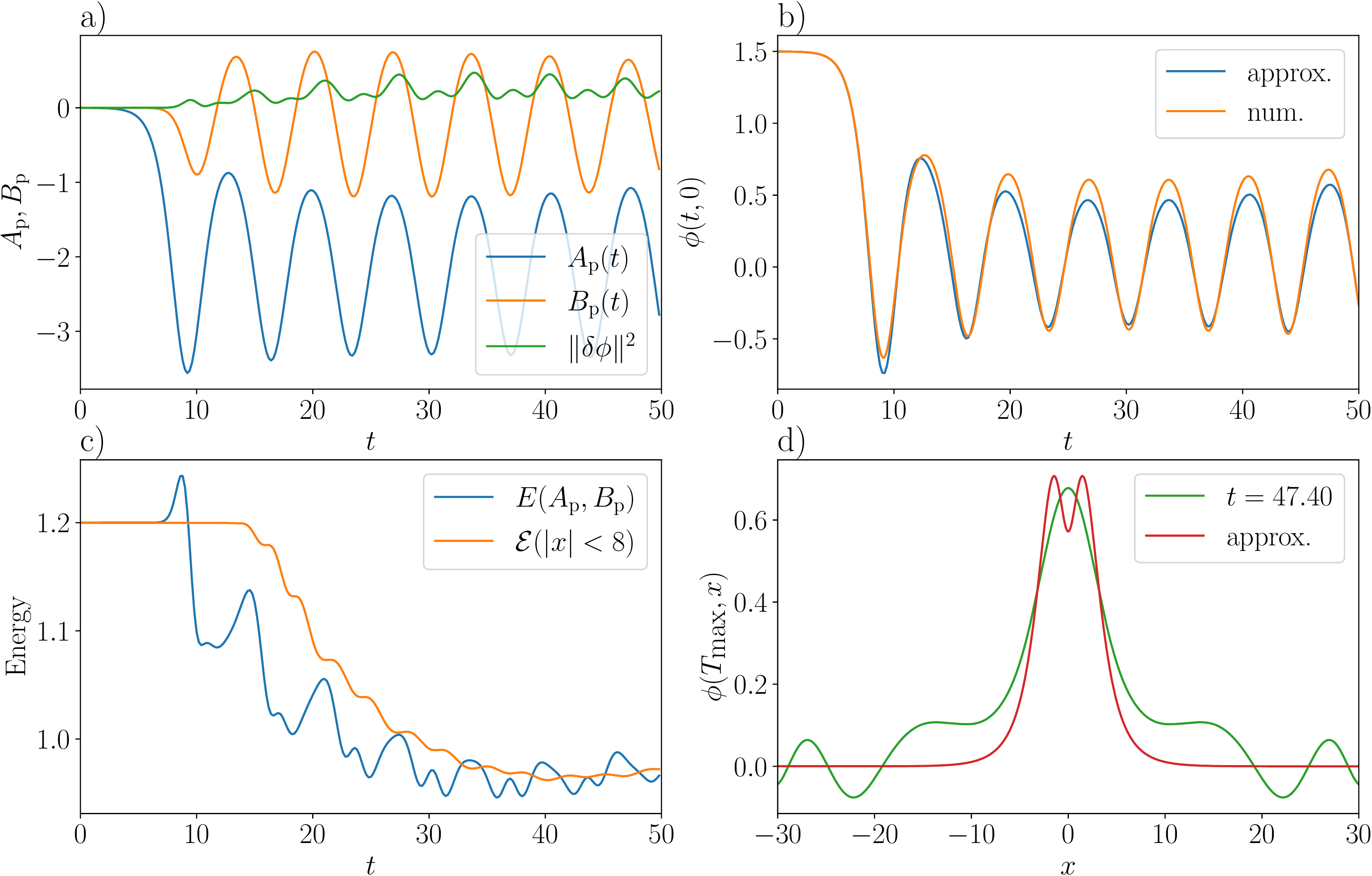}
 \caption{Evolution from initial condition
   $\phi(0,x)=\phi_{\rm S}(x)-0.001\eta_{-1}(x)$ compared with the projected
   mode parameters of the effective model.}
 \label{fig:projection1}
\end{figure}

\begin{figure}
 \includegraphics[width=0.85\columnwidth]{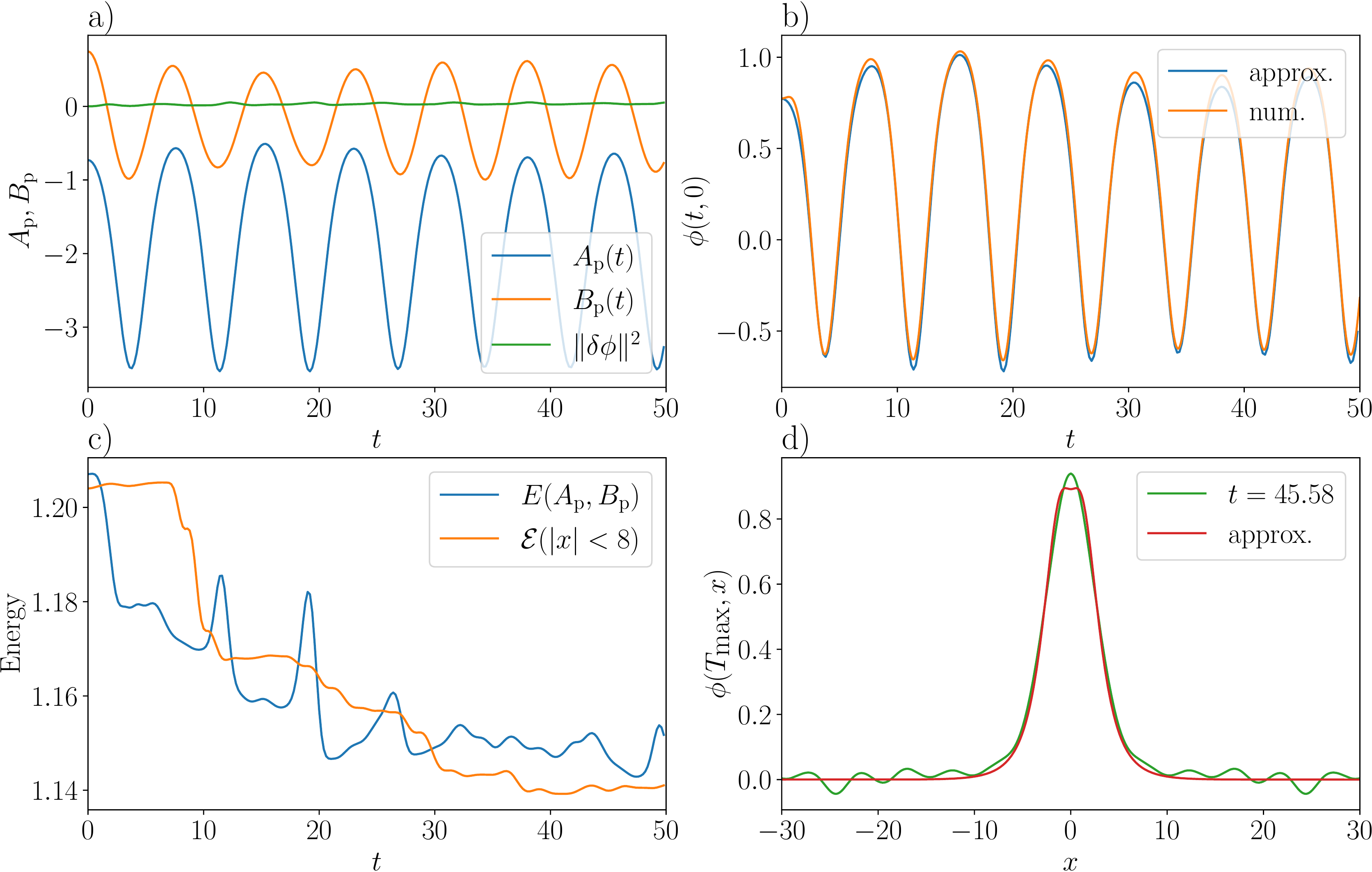}
 \caption{Evolution from initial condition
   $\phi(0,x)=\Phi_4(0,x)$ for $\epsilon=0.5$ compared
   with the projected mode parameters of the effective model.}
 \label{fig:projection2}
\end{figure}

We have explored the extent to which important features of a solution
$\phi(t,x)$ of the field equation (\ref{fieldeq}) are captured by
this reduced model. To do this it is useful to follow the amplitudes
of the projection of $\phi$ onto the modes
$\eta_{-1}$  and $\eta_1$,
\begin{equation}\label{eq:projections}
A_{\rm p}(t)=\int_{-\infty}^{\infty}
\left(\phi(t,x)-\phi_{\rm S}(x)\right)\eta_{-1}(x)\,dx \,,\qquad
B_{\rm p}(t)=\int_{-\infty}^{\infty}
\left(\phi(t,x)-\phi_{\rm S}(x)\right)\eta_{1}(x)\,dx \,.
\end{equation}
Because the modes are orthogonal to each
other and to the radiation, this method is equivalent to the
usual least squares fit of $\phi$ to the function (\ref{ansatz}), but it is
numerically faster and more stable. FIG. \ref{fig:projection1}
illustrates this approach for the numerical field evolution of a perturbed
sphaleron, with initial condition
$\phi(0,x)=\phi_{\rm S}(x)-0.001\eta_{-1}(x)$, decaying to an oscillon,
and FIG. \ref{fig:projection2} is for the evolution from the
truncated Fodor et al. series $\phi(0,x) = \Phi_4(0,x)$ with
$\epsilon=0.5$ as initial condition. The upper-left plots a)
show the values of the projected mode amplitudes $A_{\rm p}, B_{\rm p}$ and
the $L^2$ norm of the remainder $\|\delta \phi\|^2$. The upper-right
plots b) show the comparison between the field value $\phi(t,0)$ at the
centre (orange line) and its approximation $\Phi_4(t,0)$ (blue line).
The lower-left plots c) show the energy $\mathcal{E}(|x|<8)$ within
the interval $-8 < x < 8$ of the solution $\phi(t,x)$ and the energy of the
reduced model $E(A_p, B_p)$ for the fitted $A_{\rm p}(t), B_{\rm p}(t)$
values. The lower-right plots d) compare the field profiles $\phi(t,x)$
at the time $t = T_{\rm max}$ of the last maximum  of $\phi(t,0)$
before $t=50$ (green) and its projection (red).

In the case of sphaleron decay (FIG. \ref{fig:projection1}), the
oscillon is initially generated with a large amplitude, but this soon
reduces. The field is accurately approximated only until $\phi(t,0)$
crosses zero for the first time. Then the remainder grows and the
energy in the projected modes starts to decrease (unlike in the
reduced model itself). Energy starts to escape from the interval $-8 < x < 8$
in less than 20 time-units, and converts to radiation. This is
confirmed by the field profile at the latest maximum where radiation
with an amplitude approximately $0.1$ is clearly visible.

Starting from the Fodor et al. configuration $\Phi_4$ with smaller initial
amplitude (FIG. \ref{fig:projection2}), substantially less
energy is lost and all the parameters are much better approximated.
The energy decreases by about 5\% within the simulation time,
compared to over 20\% during sphaleron decay. We expect that at a
later stage of oscillon evolution, or from more carefully prepared initial
conditions, radiation would be even less. This is confirmed below.

\section{Comparison of Oscillon Models}\label{sec:comparison}

\begin{figure}
 \includegraphics[width=0.75\columnwidth]{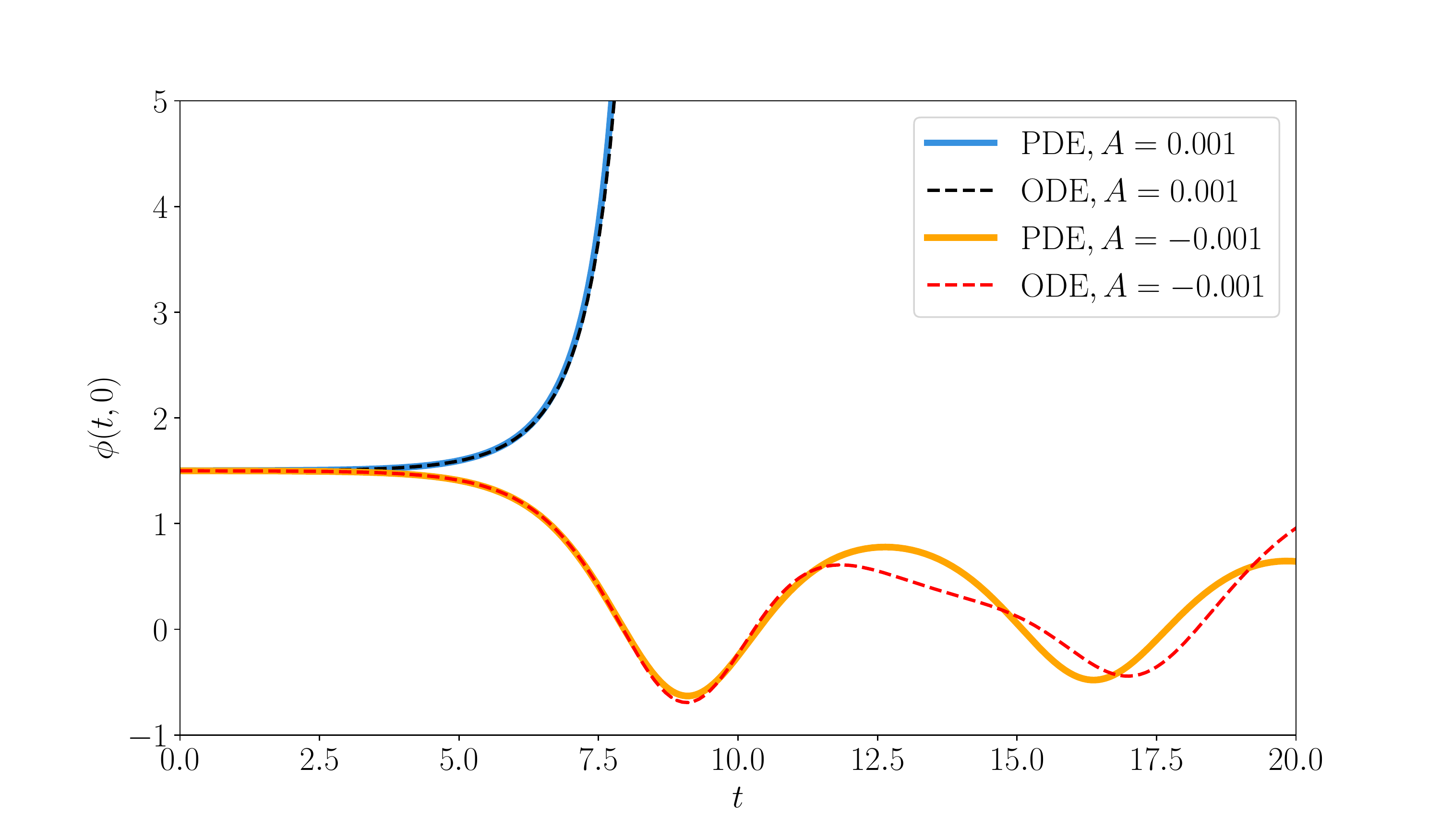}
 \caption{Solutions of the field equation (PDE) and reduced
   equations (ODEs) for initial conditions
   $\phi(0,x)=\phi_{\rm S}(x)+A \eta_{-1}(x)$.}
 \label{fig:comparison_blow}
\end{figure}

\begin{figure}
 \includegraphics[width=0.75\columnwidth]{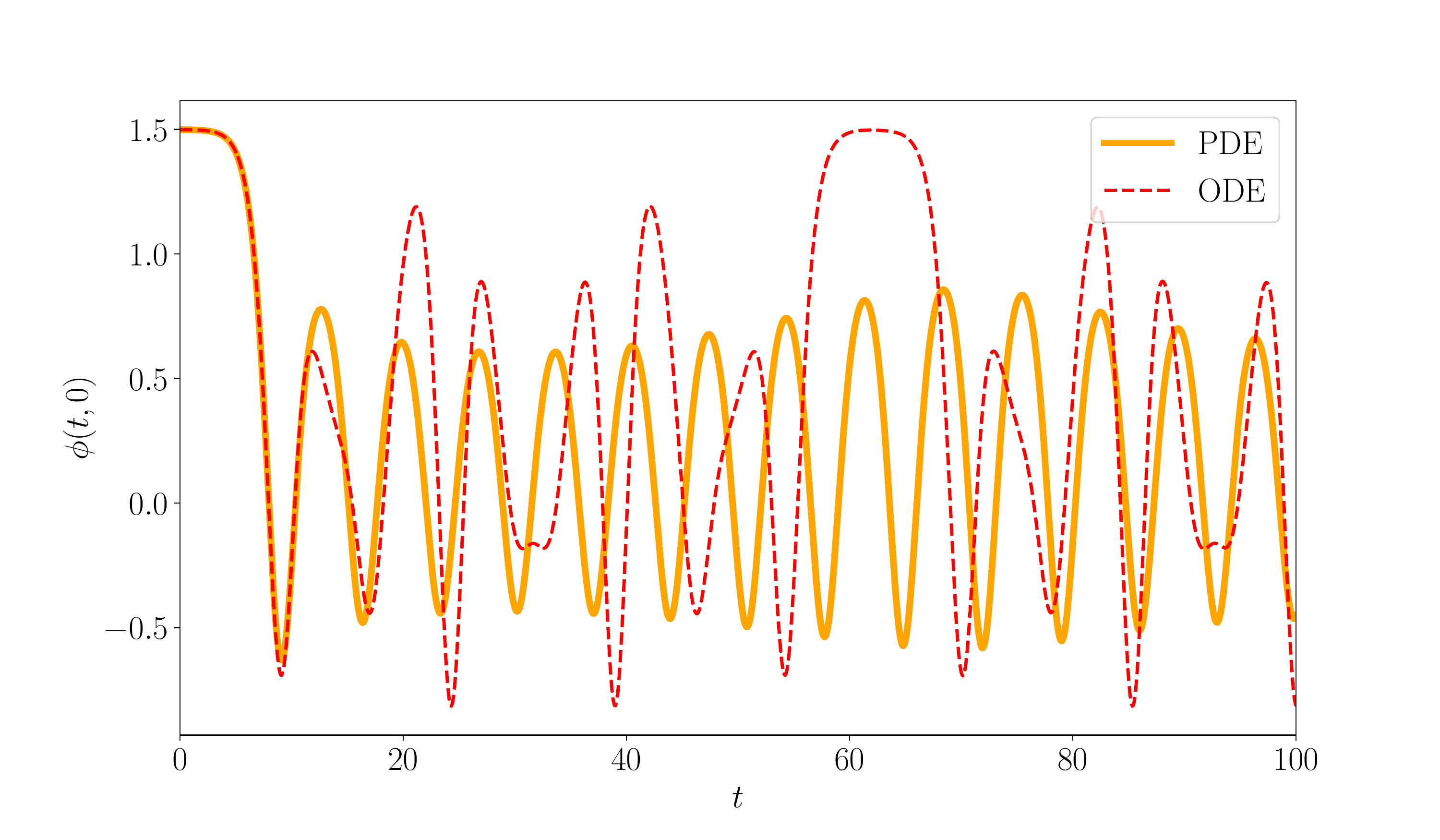}
 \caption{Longer-time solutions of the field equation (PDE) and reduced
   equations (ODEs) for initial conditions
   $\phi(0,x)=\phi_{\rm S}(x)-0.001\eta_{-1}(x)$.}
 \label{fig:oscillon_creation}
\end{figure}

\begin{figure}
 \includegraphics[width=0.75\columnwidth]{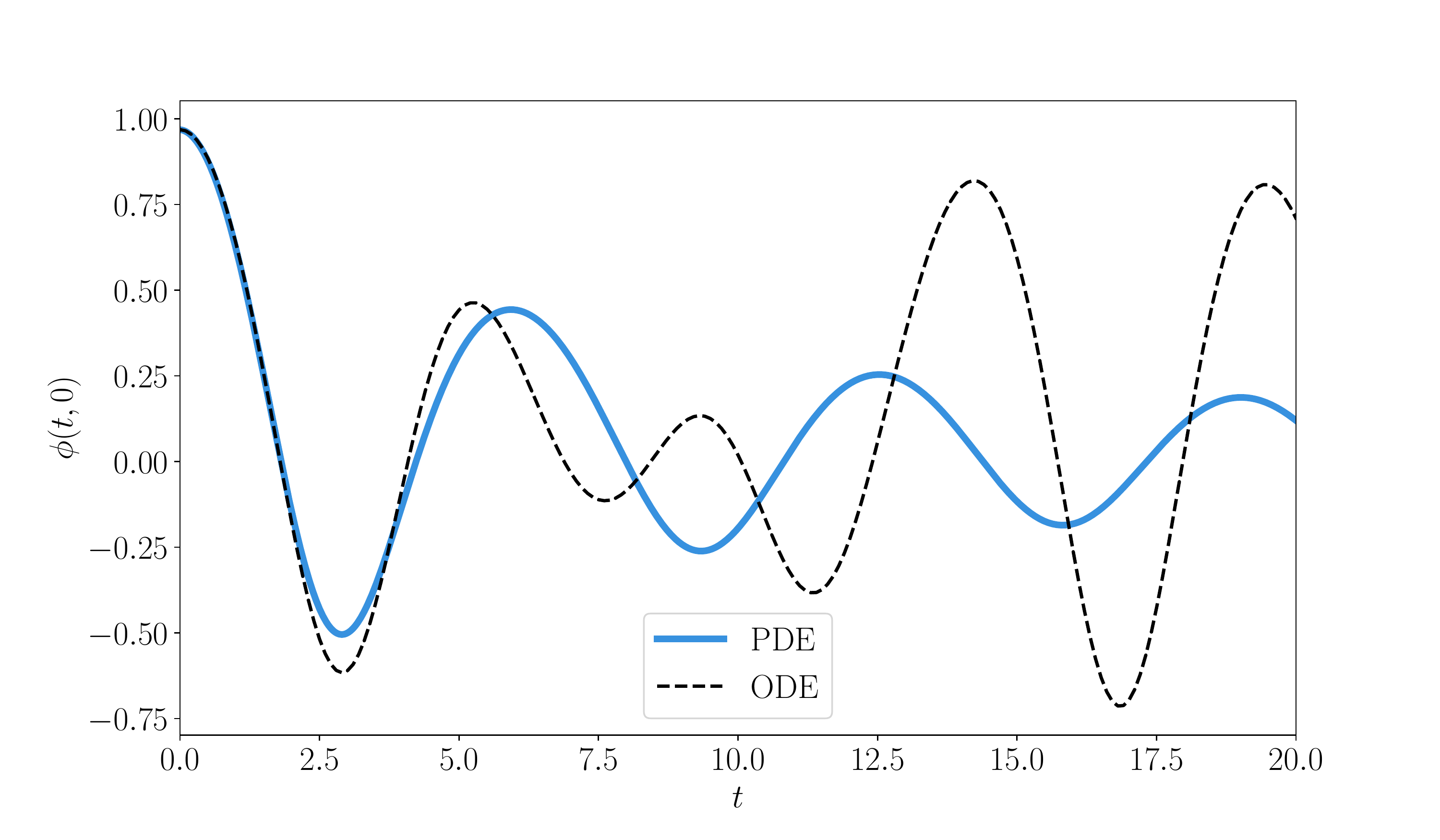}
 \caption{Solutions of the field equations (PDE) and reduced
   equations (ODEs) for initial conditions corresponding to a substantially
   deformed sphaleron with $A(0)=-1, B(0)=-0.5$.}
 \label{fig:comparison_generic}
\end{figure}

It is particularly interesting to look at the trajectory of the
reduced dynamics that starts just slightly perturbed from the
sphaleron saddle point in either direction of the unstable mode.
This is shown in FIG. \ref{fig:comparison_blow}.
A positive perturbation leads to field blow-up, and
a negative perturbation leads to oscillon formation.
In both cases, starting from the same initial conditions, the full
field dynamics (solid lines) is captured well by the reduced model
(dashed lines) in its initial stages. In the case
of oscillon formation, however, radiation
production results in a separation of trajectories.
The reduced dynamics is quasiperiodic and almost returns
to the initial amplitude $\phi(0,0)=1.499$ at $t \approx 60$.
The full field dynamics near $x=0$ becomes almost
quasiperiodic from $t \approx 15$ onwards, but it has a smaller
amplitude oscillating between 0.55 and 0.7, and a higher
basic frequency $\omega=0.9024$. Both solutions for longer times
are shown in FIG. \ref{fig:oscillon_creation}.
The difference is particularly visible for initial data for which an
oscillon does not form. An example of such evolution is shown in
FIG. \ref{fig:comparison_generic}. Again, at the initial stage, during
the first oscillation, the field and reduced dynamics are very similar.
But later the field dynamics is dominated by radiation.

\begin{figure}
 \includegraphics[width=0.85\columnwidth]{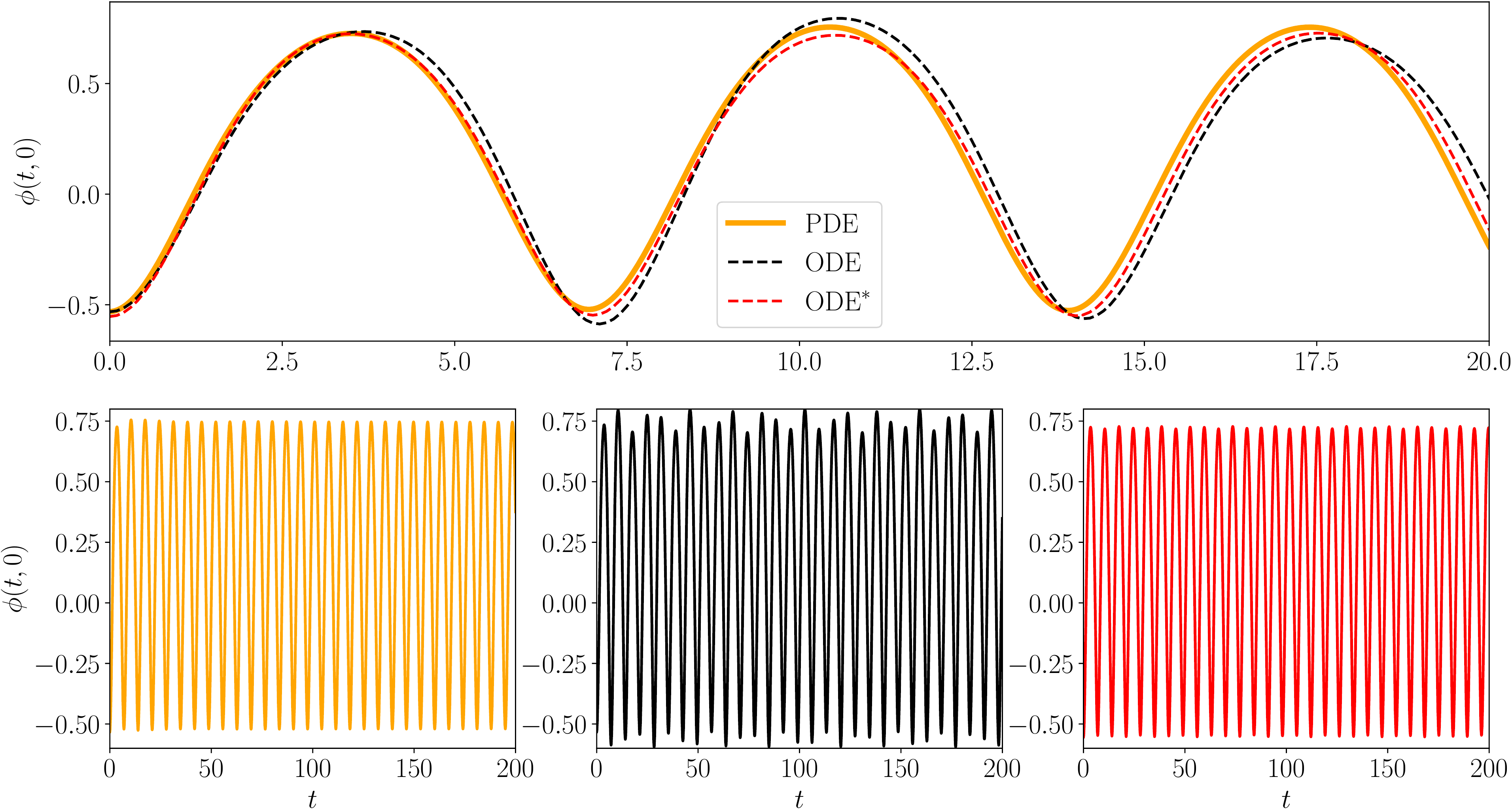}
 \caption{Solutions of the field equation (PDE) and reduced equations
   (ODEs) for initial conditions corresponding to the sphaleron modes ansatz
   with $A(0)=-3.4247$, $B(0)=-1.0218$. The red line (ODEs$^*$) is the
   periodic solution of the reduced equations with $A^*(0)=-3.4287$,
   $B^*(0)=-0.9605$.}
 \label{fig:comparison_good}
\end{figure}

The best match between the field dynamics and the dynamics of the
reduced model occurs for initial data where a true oscillon is produced with
minimal transient radiation. We have found such an oscillon for
$A(0)=-3.4247$,  $B(0)=-1.0218$. The evolution is shown in
FIG. \ref{fig:comparison_good}. The field and reduced dynamics are
very close, except for a small difference in frequencies.
However, at later times it is clear that the field and reduced dynamics do
differ. The field dynamics is almost periodic with
$\omega=0.9024$ and minimum field value $\phi_\textrm{min}=-0.5211$,
whereas the reduced dynamics is visibly quasiperiodic (bottom panel
centre) with the field minimum oscillating between $-0.59$ and $-0.53$.
This is because the field dynamics always produces some radiation,
especially early on, so the oscillon settles into a slightly different
state. But by analysing solutions of the reduced model we have found
initial conditions where the solution $\phi$ is periodic with amplitude
$\phi_\textrm{min}=-0.5485$ and frequency $\omega^*=0.8967$, very
similar to the oscillon in the field theory. This periodic solution is
also shown in FIG. \ref{fig:comparison_good}, and shown in the
$(A, B)$ plane in FIG. \ref{fig:effective_potential} (orange line).

\begin{figure}
 \includegraphics[width=0.75\columnwidth]{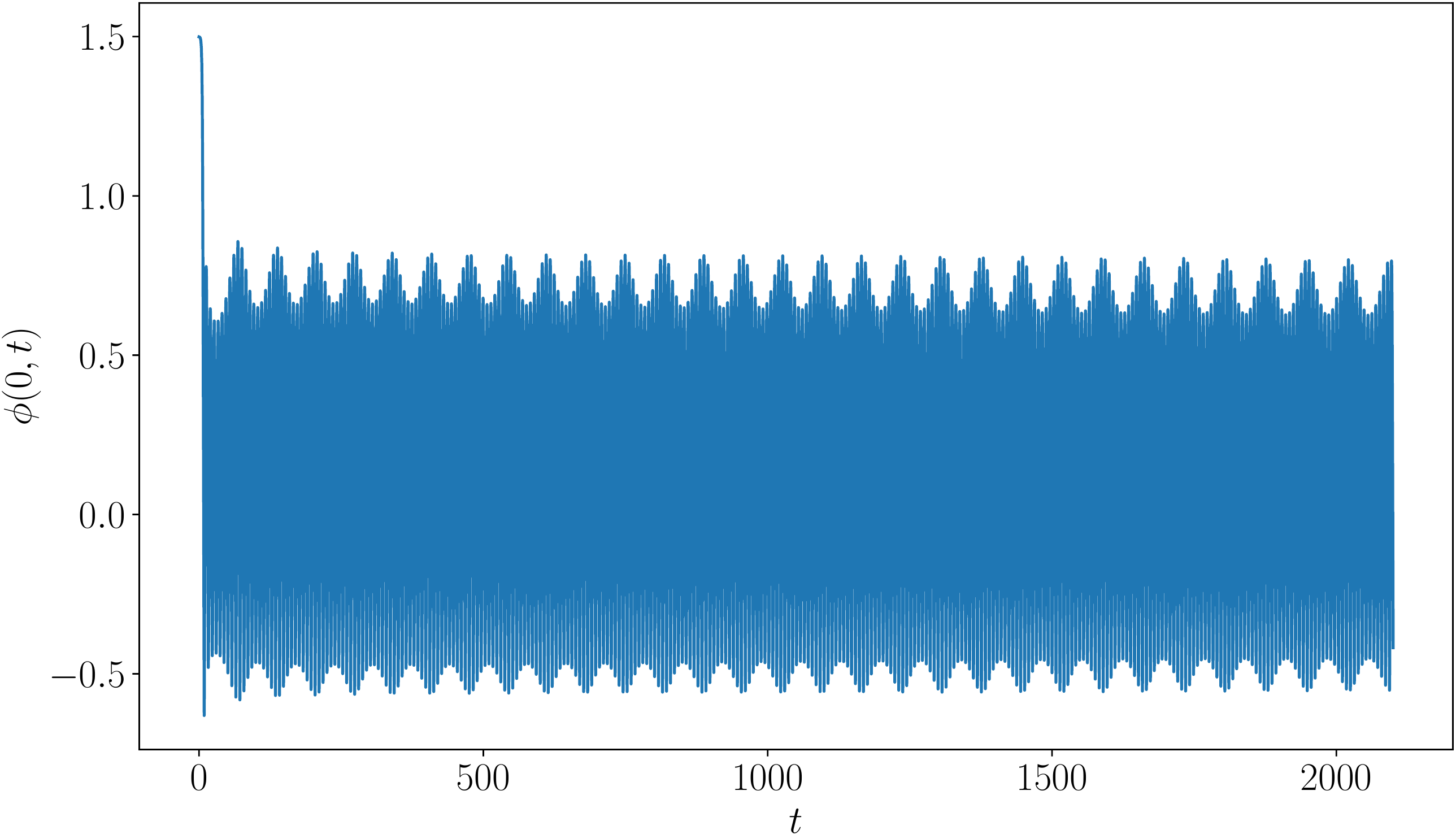}
 \caption{Approximately quasi-period oscillon arising from sphaleron
   decay with initial condition $\phi(0, x)=\phi_{\rm S}(x)
   -0.001\eta_{-1}(x)$.}\label{init_decay_longer}
\end{figure}

\begin{figure}
 \includegraphics[width=\columnwidth]{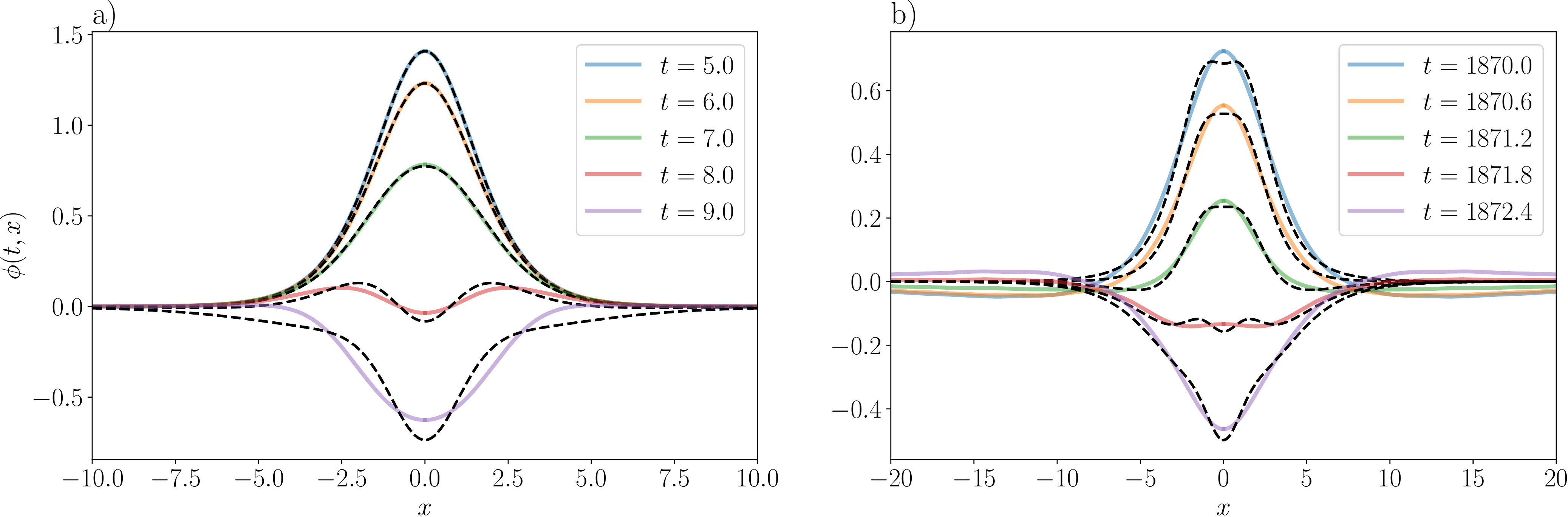}
 \caption{Decomposition of a field profile (colour lines) into
   sphaleron and its modes (dashed lines) a) at the initial stage of
   evolution and b) at a much later time.}
 \label{fig:profile_decomposition}
 \end{figure}

\begin{figure}
 \includegraphics[width=0.75\columnwidth]{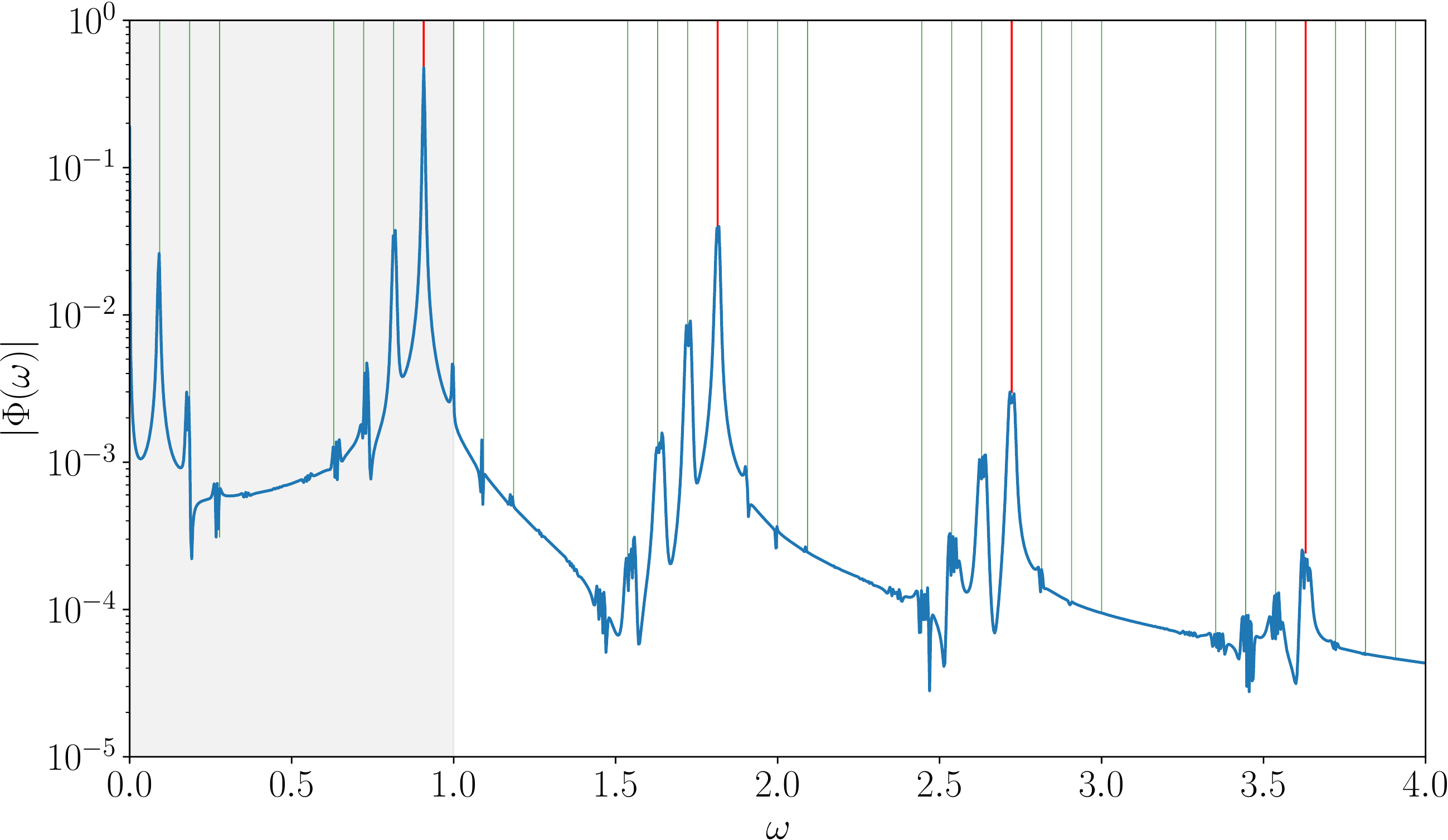}
 \caption{Power spectrum of field at center for $100<t<2000$. Red
   lines indicate harmonics of the primary frequency $\omega=0.9075$,
   the thin green lines indicate some of the combinations
   $|n_1\omega+n_2|$ for $-3\leq n_2\leq3 $.
   The widening of the peaks
   is due to a slow frequency drift.}
 \label{PowerSpectrum}
\end{figure}

By following sphaleron decay in the field theory for an even longer time,
we see that the beats persist with slightly smaller amplitude
(FIG.~\ref{init_decay_longer}). The main frequency increases to
$\omega\approx0.913$ at $t \approx 2000$. The field shape remains lump-like
but grows slightly wider (FIG.~\ref{fig:profile_decomposition}).
FIG.~\ref{PowerSpectrum} shows the power spectrum of the field
$\phi(t,x)$. It reveals the main frequency at $\omega=0.9075$ and its
harmonics. The higher harmonics are widened due to the frequency drift with
time. Around each main peak there is a family of equidistant smaller
peaks. Their positions correspond very well with $|n_1\omega+n_2|$,
which shows that there is another basic frequency near the threshold
$m=1$. These peaks are generated by resonances due to the nonlinearity
of the field equation. Recall that the reduced model
has quasi-periodic solutions but neither of the frequencies is very
close to $1$.

Not all initial conditions of the sphaleron modes ansatz
(\ref{ansatz}) give oscillons immediately, as we have seen;
one needs to minimise the energy loss to radiation, (\ref{flux}).
We have found the values of $A$ and $B$ achieving this for a range
of initial central amplitudes $\phi_0 =\phi(0, 0)$. We have also
investigated the subsequent energy loss as a function of $\phi_0$. 
For $\phi_0>0.8$ the ansatz (\ref{ansatz}) gives as good an initial condition
as the best initial condition $\Phi_2$ from the Fodor et al.
expansion. Moreover, the reduced model reproduces the amplitude
decay very well, at least up to the first minimum of the field
profile, whereas $\Phi_2$ evolves strictly periodically. The reduced
model has more degrees of freedom and has other types of
solution, one example of which will be presented in the
following section.

\begin{figure}
 \includegraphics[width=0.75\columnwidth]{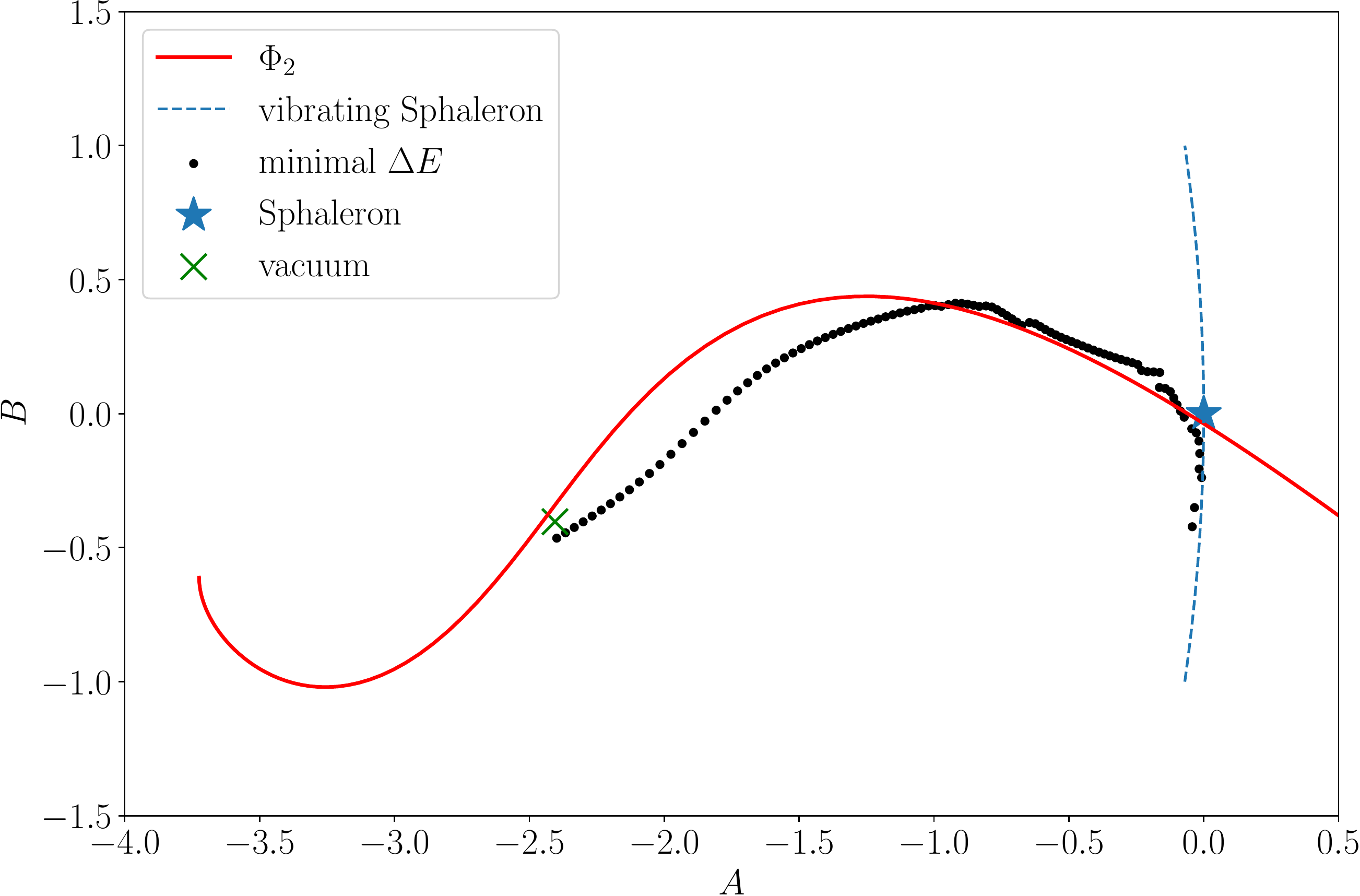}
 \caption{Comparison of the initial conditions leading to the lowest
   energy loss for given $\phi_0$ (black dots) and the $\Phi_2$
   approximation for $\epsilon\in[-0.95,0.95]$ (red line), projected
   on to the sphaleron modes. Static solutions for the
   sphaleron and approximate vacuum are also marked along with
   the solution corresponding to the vibrating sphaleron.}
 \label{fig:parametric_space}
\end{figure}

Further evidence that there is an overlap between these two
approximations is shown in FIG.~\ref{fig:parametric_space}, where on the
$(A, B)$ plane we have plotted the points corresponding to the minimal
energy loss $\Delta E$ and the projection coefficients of $\Phi_2$ on to the
modes $\eta_{-1}$ and $\eta_1$. Note that these projections do not make
much sense for small values of $\epsilon$ because the oscillon is
much wider than the sphaleron and its discrete modes. An interesting
observation can be made regarding $\Phi_2$ which lies very close to
the points of minimal energy loss. Its evolution for $\epsilon=0.728$
passes close ($A=-0.016, B=-0.027$) to the sphaleron with $A=B=0$.
Some points obtained by minimising the energy loss also lie very close
to the line corresponding to the vibrating sphaleron, which we discuss next.

\section{Vibrating Sphaleron}

At the linearised level of the reduced model, with Lagrangian
(\ref{Lageff}), the shape mode $\eta_1$ of the sphaleron oscillates
indefinitely with constant amplitude ${\cal B}$ and
frequency $\omega_1 = \sqrt{3/4}$. However, the nonlinear coupling
of $B$ to $A$ leads to an excitation of the unstable mode $\eta_{-1}$
as well. In eq.(\ref{EffectiveEqsa}), the term $C_3B^2$ can cause exponential
growth of $A$. We see this in more detail by solving eqs.(\ref{EffectiveEqsa},
\ref{EffectiveEqsb}) to low order in ${\cal B}$. Assume that at
linear order only the shape mode is excited, so
\begin{equation}
 A(t)=0, \qquad B(t)={\cal B}\cos(\omega_1 t) \,.
\end{equation}
At quadratic order,
\begin{equation}
 \frac{d^2 A}{dt^2} = \frac{5}{4} A + C_3 B^2 = \frac{5}{4} A
 + \frac{1}{2}C_3 {\cal B}^2 + \frac{1}{2}C_3 {\cal B}^2 \cos(2\omega_1 t) \,,
 \end{equation}
whose general solution is
\begin{equation}
A(t) = F_1e^{|\omega_{-1}|t} + F_2e^{-|\omega_{-1}|t} -
C_3{\cal B}^2\left(\frac{2}{5}+\frac{2}{5+16\omega_1^2}
\cos(2\omega_1t)\right) \,,
\end{equation}
where $|\omega_{-1}| = \sqrt{5/4}$.

\begin{figure}
 \includegraphics[width=0.75\columnwidth]{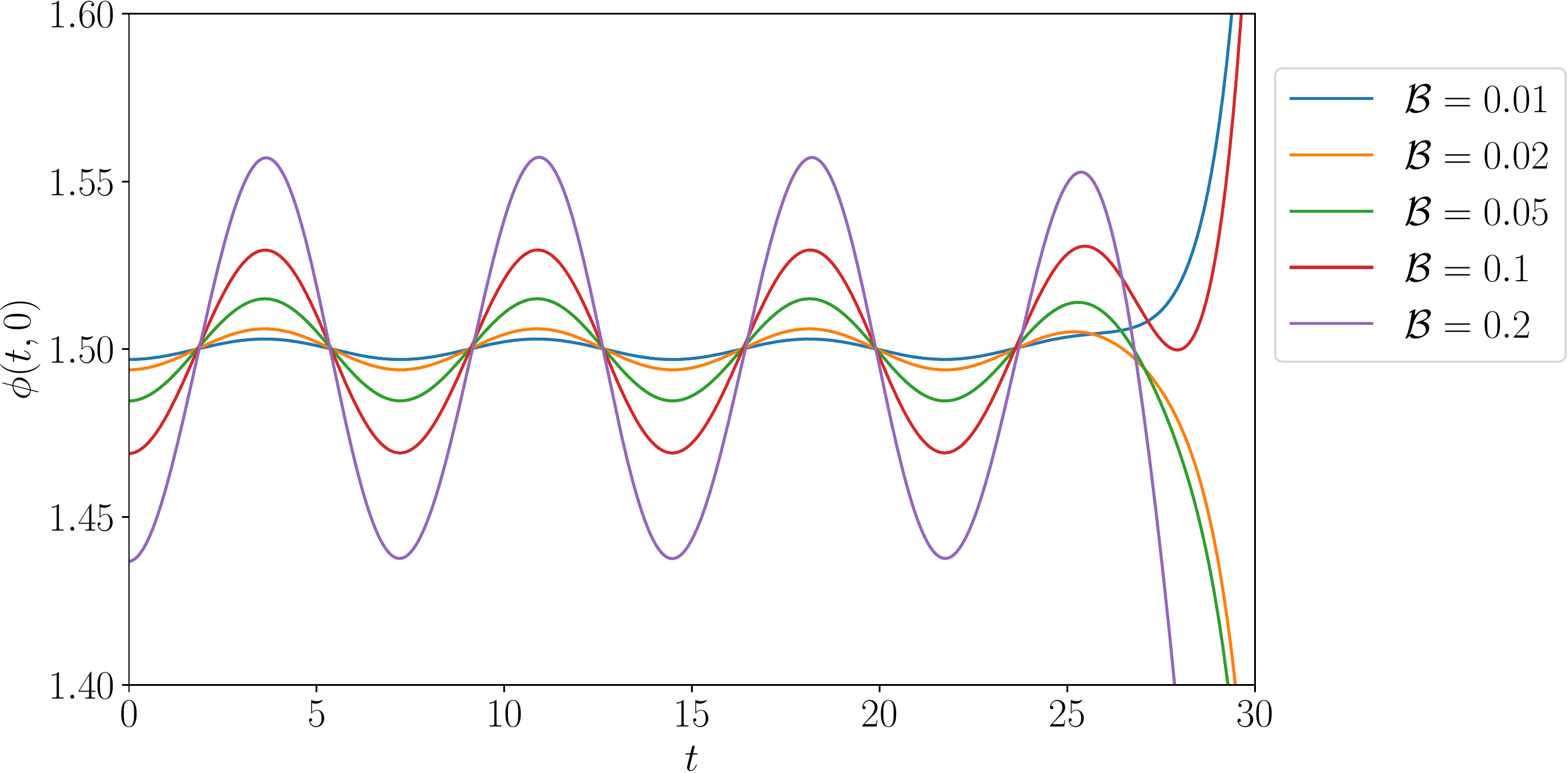}
 \caption{Nearly periodic evolution of the vibrating sphaleron
   with fine-tuned initial data.}
 \label{fig:vibr_sphaleron}
\end{figure}

Generally, $F_1$ is non-zero and the solution grows exponentially.
However, for the initial conditions
\begin{equation}
  \frac{dA}{dt}(0)=0,\qquad A(0)=
  -C_3{\cal B}^2\left(\frac{2}{5}+\frac{2}{5+16\omega_1^2}\right)
  =-\frac{44}{85}C_3{\cal B}^2
\end{equation}
there are no exponential terms in the solution, and only constant
and oscillatory terms remain. Motivated by this, we have fine-tuned the
initials conditions in the field theory and obtained an almost
periodic, vibrating sphaleron with a varying amplitude of oscillation,
as shown in FIG. \ref{fig:vibr_sphaleron}. For initial conditions we took
\begin{equation}
  \phi_t(0,x)=0 \,, \qquad
  \phi(0,x)=\phi_{\rm S}(x) - \alpha\frac{44}{85}C_3{\cal B}^2\eta_{-1}(x) +
 {\cal B}\eta_1(x) \,,
 \end{equation}
where $\alpha$ was chosen to suppress exponential growth for as long
as possible. $\alpha = 1.024$ for ${\cal B}=0.2$ and $\alpha$
decreases to 1 as ${\cal B}$ approaches 0. Similar vibrating sphaleron
solutions were considered earlier at linear order in \cite{Bizon:2011zz}.

\section{Conclusions}

We have considered a simple 1-dimensional scalar field theory with a
cubic potential, having a long-lived oscillon solution as well as a
static, unstable sphaleron solution. We have explicitly constructed
the Fodor et al. expansion for the oscillon up to fourth order
in the oscillon's amplitude parameter. As this 
parameter increases, the oscillon frequency decreases. The expansion is
asymptotic rather than convergent, so the fourth-order truncation
$\Phi_4(t,x)$ is only valid for small amplitudes. A larger-amplitude
oscillon is better approximated by the second-order truncation
$\Phi_2(t,x)$.

When the oscillon is instantaneously at rest, its shape is similar to
that of the sphaleron, but the sphaleron has larger amplitude and
more energy. The sphaleron, slightly perturbed, decays into the
oscillon. During its first couple of oscillations it radiates a significant
fraction of its energy, but then settles into an oscillon of
relatively large amplitude.

The Fodor et al. oscillon is periodic, with a single fundamental
frequency. However, the decaying sphaleron approaches an oscillon
whose amplitude is itself slightly oscillating. This suggests that
oscillon solutions are best modelled by a truncation of the field
theory having two degrees of freedom. The sphaleron naturally provides
these -- it has a single unstable mode and a further discrete mode
of oscillation whose frequency is below the threshold frequency for the
continuum of radiation modes.

We have considered the field ansatz
obtained by linearly deforming the sphaleron by these two modes, with
amplitudes $A$ and $B$. Substituting this ansatz into the full
field-theory Lagrangian, we obtain a reduced, nonlinear dynamical Lagrangian
for $A$ and $B$, whose nonlinearity arises from the cubic potential
term. This dynamics is decoupled from the
field radiation modes. Because the ansatz gets quite close to the
vacuum configuration for particular values of $A$ and $B$, it provides
a useful interpolation between the sphaleron and the vacuum. The dynamical
equations for $A$ and $B$ have solutions describing sphaleron decay as
well as oscillons of relatively large amplitude. However,
oscillons of small amplitude, which have a larger spatial extent than
the sphaleron and its discrete modes, are not well-described by the ansatz.

By carefully adjusting the initial conditions of $A$ and $B$, we can
find oscillatory solutions of the reduced dynamics that are almost
exactly periodic. We can also use these initial conditions as initial
conditions for the field theory itself, to generate oscillons with
minimal radiation. We then find close similarities between the field
theory dynamics and the reduced dynamics of $A$ and $B$. The
comparison is effected by projecting the field dynamics on to the
two discrete modes.

In summary, we have found similar oscillon solutions from several points of
view -- through the Fodor et al. small-amplitude expansion, through the decay
of the sphaleron and oscillating versions of the sphaleron, and
from our truncation of the field theory to a dynamical system with two
degrees of freedom. This dynamical system appears to be a useful
extension of the truncation to one degree of freedom implied by the
Fodor et al. analysis. However, all these approaches are only
approximate. The Fodor et al. expansion is asymptotic and needs to
be truncated; the decaying sphaleron emits considerable radiation in
its initial few oscilations; finally, our linearised field ansatz
exploiting the sphaleron's discrete modes misses the vacuum and
small-amplitude oscillons.

Numerically, there is overwhelming evidence for the existence of
long-lived, topologically-trivial, localised oscillatory solutions of
the field theory -- oscillons -- but the precise mathematical status
of oscillons remains elusive.

\section*{Acknowledgements}

The research of TR was supported by the Polish National Science Centre,
grant number NCN 2019/35/B/ST2/00059.

\bibliographystyle{apsrev4-2}
\bibliography{refs}

%apsrev4-2.bst 2019-01-14 (MD) hand-edited version of apsrev4-1.bst
%Control: key (0)
%Control: author (72) initials jnrlst
%Control: editor formatted (1) identically to author
%Control: production of article title (-1) disabled
%Control: page (0) single
%Control: year (1) truncated
%Control: production of eprint (0) enabled
\begin{thebibliography}{10}%
\makeatletter
\providecommand \@ifxundefined [1]{%
 \@ifx{#1\undefined}
}%
\providecommand \@ifnum [1]{%
 \ifnum #1\expandafter \@firstoftwo
 \else \expandafter \@secondoftwo
 \fi
}%
\providecommand \@ifx [1]{%
 \ifx #1\expandafter \@firstoftwo
 \else \expandafter \@secondoftwo
 \fi
}%
\providecommand \natexlab [1]{#1}%
\providecommand \enquote  [1]{``#1''}%
\providecommand \bibnamefont  [1]{#1}%
\providecommand \bibfnamefont [1]{#1}%
\providecommand \citenamefont [1]{#1}%
\providecommand \href@noop [0]{\@secondoftwo}%
\providecommand \href [0]{\begingroup \@sanitize@url \@href}%
\providecommand \@href[1]{\@@startlink{#1}\@@href}%
\providecommand \@@href[1]{\endgroup#1\@@endlink}%
\providecommand \@sanitize@url [0]{\catcode `\\12\catcode `\$12\catcode
  `\&12\catcode `\#12\catcode `\^12\catcode `\_12\catcode `\%12\relax}%
\providecommand \@@startlink[1]{}%
\providecommand \@@endlink[0]{}%
\providecommand \url  [0]{\begingroup\@sanitize@url \@url }%
\providecommand \@url [1]{\endgroup\@href {#1}{\urlprefix }}%
\providecommand \urlprefix  [0]{URL }%
\providecommand \Eprint [0]{\href }%
\providecommand \doibase [0]{https://doi.org/}%
\providecommand \selectlanguage [0]{\@gobble}%
\providecommand \bibinfo  [0]{\@secondoftwo}%
\providecommand \bibfield  [0]{\@secondoftwo}%
\providecommand \translation [1]{[#1]}%
\providecommand \BibitemOpen [0]{}%
\providecommand \bibitemStop [0]{}%
\providecommand \bibitemNoStop [0]{.\EOS\space}%
\providecommand \EOS [0]{\spacefactor3000\relax}%
\providecommand \BibitemShut  [1]{\csname bibitem#1\endcsname}%
\let\auto@bib@innerbib\@empty
%</preamble>
\bibitem [{\citenamefont {Gleiser}\ and\ \citenamefont
  {Sicilia}(2009)}]{GleSic}%
  \BibitemOpen
  \bibfield  {author} {\bibinfo {author} {\bibfnamefont {M.}~\bibnamefont
  {Gleiser}}\ and\ \bibinfo {author} {\bibfnamefont {D.}~\bibnamefont
  {Sicilia}},\ }\href {https://doi.org/10.1103/PhysRevD.80.125037} {\bibfield
  {journal} {\bibinfo  {journal} {Phys. Rev. D}\ }\textbf {\bibinfo {volume}
  {80}},\ \bibinfo {pages} {125037} (\bibinfo {year} {2009})},\ \Eprint
  {https://arxiv.org/abs/0910.5922} {arXiv:0910.5922 [hep-th]} \BibitemShut
  {NoStop}%
\bibitem [{\citenamefont {Klinkhamer}\ and\ \citenamefont
  {Manton}(1984)}]{KliMan}%
  \BibitemOpen
  \bibfield  {author} {\bibinfo {author} {\bibfnamefont {F.~R.}\ \bibnamefont
  {Klinkhamer}}\ and\ \bibinfo {author} {\bibfnamefont {N.~S.}\ \bibnamefont
  {Manton}},\ }\href {https://doi.org/10.1103/PhysRevD.30.2212} {\bibfield
  {journal} {\bibinfo  {journal} {Phys. Rev. D}\ }\textbf {\bibinfo {volume}
  {30}},\ \bibinfo {pages} {2212} (\bibinfo {year} {1984})}\BibitemShut
  {NoStop}%
\bibitem [{\citenamefont {Dorey}\ \emph {et~al.}(2020)\citenamefont {Dorey},
  \citenamefont {Roma\'{n}czukiewicz},\ and\ \citenamefont
  {Shnir}}]{Dorey:2019uap}%
  \BibitemOpen
  \bibfield  {author} {\bibinfo {author} {\bibfnamefont {P.}~\bibnamefont
  {Dorey}}, \bibinfo {author} {\bibfnamefont {T.}~\bibnamefont
  {Roma\'{n}czukiewicz}},\ and\ \bibinfo {author} {\bibfnamefont
  {Y.}~\bibnamefont {Shnir}},\ }\href
  {https://doi.org/10.1016/j.physletb.2020.135497} {\bibfield  {journal}
  {\bibinfo  {journal} {Phys. Lett. B}\ }\textbf {\bibinfo {volume} {806}},\
  \bibinfo {pages} {135497} (\bibinfo {year} {2020})},\ \Eprint
  {https://arxiv.org/abs/1910.04128} {arXiv:1910.04128 [hep-th]} \BibitemShut
  {NoStop}%
\bibitem [{\citenamefont {Fodor}\ \emph {et~al.}(2008)\citenamefont {Fodor},
  \citenamefont {Forg\'{a}cs}, \citenamefont {Horv\'{a}th},\ and\ \citenamefont
  {Luk\'{a}cs}}]{Fodor:2008es}%
  \BibitemOpen
  \bibfield  {author} {\bibinfo {author} {\bibfnamefont {G.}~\bibnamefont
  {Fodor}}, \bibinfo {author} {\bibfnamefont {P.}~\bibnamefont {Forg\'{a}cs}},
  \bibinfo {author} {\bibfnamefont {Z.}~\bibnamefont {Horv\'{a}th}},\ and\
  \bibinfo {author} {\bibfnamefont {A.}~\bibnamefont {Luk\'{a}cs}},\ }\href
  {https://doi.org/10.1103/PhysRevD.78.025003} {\bibfield  {journal} {\bibinfo
  {journal} {Phys. Rev. D}\ }\textbf {\bibinfo {volume} {78}},\ \bibinfo
  {pages} {025003} (\bibinfo {year} {2008})},\ \Eprint
  {https://arxiv.org/abs/0802.3525} {arXiv:0802.3525 [hep-th]} \BibitemShut
  {NoStop}%
\bibitem [{\citenamefont {Perring}\ and\ \citenamefont
  {Skyrme}(1962)}]{PerSky}%
  \BibitemOpen
  \bibfield  {author} {\bibinfo {author} {\bibfnamefont {J.~K.}\ \bibnamefont
  {Perring}}\ and\ \bibinfo {author} {\bibfnamefont {T.~H.~R.}\ \bibnamefont
  {Skyrme}},\ }\href
  {https://doi.org/https://doi.org/10.1016/0029-5582(62)90774-5} {\bibfield
  {journal} {\bibinfo  {journal} {Nucl. Phys.}\ }\textbf {\bibinfo {volume}
  {31}},\ \bibinfo {pages} {550} (\bibinfo {year} {1962})}\BibitemShut
  {NoStop}%
\bibitem [{\citenamefont {Dorey}\ \emph {et~al.}()\citenamefont {Dorey},
  \citenamefont {Gorina}, \citenamefont {Perapechka}, \citenamefont
  {Roma\'nczukiewicz},\ and\ \citenamefont {Shnir}}]{Dorey:2021mdh}%
  \BibitemOpen
  \bibfield  {author} {\bibinfo {author} {\bibfnamefont {P.}~\bibnamefont
  {Dorey}}, \bibinfo {author} {\bibfnamefont {A.}~\bibnamefont {Gorina}},
  \bibinfo {author} {\bibfnamefont {I.}~\bibnamefont {Perapechka}}, \bibinfo
  {author} {\bibfnamefont {T.}~\bibnamefont {Roma\'nczukiewicz}},\ and\
  \bibinfo {author} {\bibfnamefont {Y.}~\bibnamefont {Shnir}},\ }\href
  {https://doi.org/10.1007/JHEP09(2021)145} {\bibfield  {journal} {\bibinfo
  {journal} {JHEP}\ }\textbf {\bibinfo {volume} {09}},\ \bibinfo {pages} {145
  (2021)}},\ \Eprint {https://arxiv.org/abs/2106.09560} {arXiv:2106.09560
  [hep-th]} \BibitemShut {NoStop}%
\bibitem [{\citenamefont {Callan}\ and\ \citenamefont
  {Coleman}(1977)}]{CallanCurtis:1977}%
  \BibitemOpen
  \bibfield  {author} {\bibinfo {author} {\bibfnamefont {C.~G.}\ \bibnamefont
  {Callan}, \bibfnamefont {Jr.}}\ and\ \bibinfo {author} {\bibfnamefont
  {S.}~\bibnamefont {Coleman}},\ }\href
  {https://doi.org/10.1103/PhysRevD.16.1762} {\bibfield  {journal} {\bibinfo
  {journal} {Phys. Rev. D}\ }\textbf {\bibinfo {volume} {16}},\ \bibinfo
  {pages} {1762} (\bibinfo {year} {1977})}\BibitemShut {NoStop}%
\bibitem [{\citenamefont {Avelar}\ \emph {et~al.}(2008)\citenamefont {Avelar},
  \citenamefont {Bazeia}, \citenamefont {Losano},\ and\ \citenamefont
  {Menezes}}]{Avelar:2008}%
  \BibitemOpen
  \bibfield  {author} {\bibinfo {author} {\bibfnamefont {A.~T.}\ \bibnamefont
  {Avelar}}, \bibinfo {author} {\bibfnamefont {D.}~\bibnamefont {Bazeia}},
  \bibinfo {author} {\bibfnamefont {L.}~\bibnamefont {Losano}},\ and\ \bibinfo
  {author} {\bibfnamefont {R.}~\bibnamefont {Menezes}},\ }\href
  {https://doi.org/10.1140/epjc/s10052-008-0578-6} {\bibfield  {journal}
  {\bibinfo  {journal} {Eur. Phys. J. C}\ }\textbf {\bibinfo {volume} {55}},\
  \bibinfo {pages} {133} (\bibinfo {year} {2008})},\ \Eprint
  {https://arxiv.org/abs/0711.4721} {arXiv:0711.4721 [hep-th]} \BibitemShut
  {NoStop}%
\bibitem [{\citenamefont {Blaschke}\ and\ \citenamefont
  {Karp\'\i{}\v{s}ek}(2022)}]{Blaschke:2022fxp}%
  \BibitemOpen
  \bibfield  {author} {\bibinfo {author} {\bibfnamefont {F.}~\bibnamefont
  {Blaschke}}\ and\ \bibinfo {author} {\bibfnamefont {O.~N.}\ \bibnamefont
  {Karp\'\i{}\v{s}ek}},\ }\href {https://doi.org/10.1093/ptep/ptac104}
  {\bibfield  {journal} {\bibinfo  {journal} {PTEP}\ }\textbf {\bibinfo
  {volume} {2022}},\ \bibinfo {pages} {103A01} (\bibinfo {year} {2022})},\
  \Eprint {https://arxiv.org/abs/2202.05675} {arXiv:2202.05675 [hep-th]}
  \BibitemShut {NoStop}%
\bibitem [{\citenamefont {Bizo\'{n}}\ \emph {et~al.}(2011)\citenamefont
  {Bizo\'{n}}, \citenamefont {Chmaj},\ and\ \citenamefont
  {Szpak}}]{Bizon:2011zz}%
  \BibitemOpen
  \bibfield  {author} {\bibinfo {author} {\bibfnamefont {P.}~\bibnamefont
  {Bizo\'{n}}}, \bibinfo {author} {\bibfnamefont {T.}~\bibnamefont {Chmaj}},\
  and\ \bibinfo {author} {\bibfnamefont {N.}~\bibnamefont {Szpak}},\ }\href
  {https://doi.org/10.1063/1.3645363} {\bibfield  {journal} {\bibinfo
  {journal} {J. Math. Phys.}\ }\textbf {\bibinfo {volume} {52}},\ \bibinfo
  {pages} {103703} (\bibinfo {year} {2011})},\ \Eprint
  {https://arxiv.org/abs/1012.1033} {arXiv:1012.1033 [math-ph]} \BibitemShut
  {NoStop}%
\end{thebibliography}%

\end{document}